\documentclass[twocolumn,tighten]{aastex63}

\usepackage{amsmath}
\usepackage{booktabs}
\usepackage{bbding}

\newcommand{\unit}[1]{\ensuremath{\,\mathrm{#1}}}
\newcommand{\ie}{\textit{i.e.}, }

\newcommand{\rsat}{{\rm sat}}

\newcommand{\rbaryon}{{\rm baryon}}
\newcommand{\rhole}{{\rm hole}}
\newcommand{\ramb}{{\rm amb}}
\newcommand{\rgrav}{{\rm grav}}

\newcommand{\rmax}{{\rm max}}
\newcommand{\rBH}{{\rm BH}}
\newcommand{\reff}{{\rm eff}}
\newcommand{\rkin}{{\rm kin}}
\newcommand{\rtot}{{\rm tot}}
\newcommand{\rejecta}{{\rm ejecta}}

\defcitealias{fernandez:18}{F18}
\defcitealias{ivanov:21}{I21}
\defcitealias{schneider:20}{S20}

\submitjournal{ApJ}

\shorttitle{A Parameterized Model to Study Failed CCSNe}
\shortauthors{da Silva Schneider \& O'Connor}

\begin{document}


\title{A Parameterized Neutrino Emission Model to Study Mass Ejection in Failed Core-collapse Supernovae}

\correspondingauthor{Andr\'{e} da Silva Schneider}
\email{andre.schneider@astro.su.se}

\author[0000-0003-0849-7691]{Andr\'{e} da Silva Schneider}
\affil{The Oskar Klein Centre, Department of Astronomy,
Stockholm University, AlbaNova, SE-106 91 
Stockholm, Sweden}

\author[0000-0002-8228-796X]{Evan O'Connor}
\affil{The Oskar Klein Centre, Department of Astronomy,
Stockholm University, AlbaNova, SE-106 91 
Stockholm, Sweden}


\date{\today}

\begin{abstract}

Some massive stars end their lives as \textit{failed} core-collapse supernovae (CCSNe) and become black holes (BHs). 
Although in this class of phenomena the stalled supernova shock is not revived, the outer stellar envelope can still be partially ejected. 
This occurs because the hydrodynamic equilibrium of the star is disrupted by the gravitational mass loss of the protoneutron star (PNS) due to neutrino emission. 
We develop a simple parameterized model that emulates PNS evolution and its neutrino emission and use it to simulate failed CCSNe in spherical symmetry for a wide range of progenitor stars. 
Our model allows us to study mass ejection of failed CCSNe where the PNS collapses into a BH within $\sim100\,{\rm ms}$ and up to $\sim10^6\,{\rm s}$. 
We perform failed CCSNe simulations for 262 different pre-SN progenitors and determine how the energy and mass of the ejecta depend on progenitor properties and the equation of state (EOS) of dense matter. 
In the case of a future failed CCSN observation, the trends obtained in our simulations can be used to place constraints on the pre-SN progenitor characteristics, the EOS, and on PNS properties at BH formation time. 

\end{abstract}


\section{Introduction}

Massive stars with zero-age main-sequence (ZAMS) mass $M_{\rm ZAMS} \gtrsim 8\,M_\odot$ undergo a core-collapse event once their nuclear fuel is exhausted. 
As the inert iron-core collapses onto itself, its density and pressure increase by orders of magnitude within a fraction of a second. 
This occurs until densities reach nuclear saturation density, $\rho_{\rm sat}\simeq 2.7\times10^{14} \unit{g\,cm}^{-3}$, and collapse is suddenly halted by the stiffening of nuclear interactions. 
At this moment, a protoneutron star (PNS) is born where once was the iron core of the massive star. 
Meanwhile, continuously infalling matter hits the hard PNS and rebounds, creating a shock wave that propagates outwards through the still accreting outer layers of the star. 
Eventually this shock stalls, due to photodissociation of heavy nuclei and neutrino emission losses, and starts to recede. 
As matter continues to accrete onto the PNS, neutrino emission increases. 
If neutrinos leaving the PNS deposit enough energy behind the stalled shock, the shock is revived, unbinding most of the outer layers of the star and igniting a bright \textit{successful} core-collapse supernova (CCSN).

However, a still unknown fraction of massive stars are expected to undergo a \textit{failed} CCSN event, \ie the shock wave is not revived and recedes as matter continues to accrete onto the PNS. 
Once enough matter is accreted, the PNS overcomes the maximum mass it can support and collapses into a black hole (BH) without a bright transient. 
This BH formation channel could occur for some massive stars and determining the landscape of which massive stars lead to successful explosions and which ones do not is an active topic of research \citep{sukhbold:16, couch:20, burrows:20, boccioli:22a}. 
Despite BH formation, mass ejection is still possible for failed SNe because of the hydrodynamic consequences of the neutrino emission that carries away $\mathcal{O}(10\%)$ of the gravitational mass of the PNS. 
Although these neutrinos barely interact with the outer regions of the star as they fly by, their emission decreases the gravitational pull of the PNS on the outer layers of the collapsing star. 
The sudden disruption of the hydrostatic equilibrium creates a pressure wave deep within the star that travels outwards, gaining or losing momentum depending on local stellar properties. 
If the pulse gains enough momentum it eventually forms a shock that becomes unbound and escapes to the interstellar medium \citep{nadyozhin:80, lovegrove:13, fernandez:18}. 
In such a BH formation channel, the progenitor star would become momentarily brighter, for a few seconds up to a few days, before dimming significantly or even disappear altogether. 
Furthermore, ejection of the outer stellar envelope could lead to a bright transient, even if it is orders of magnitude less luminous than typical successful SN. 
Although a failed SN has never been directly observed, there is some evidence that they do occur \citep{gerke:15, adams:17, allan:20, murguia-berthier:20, basinger:21, neustadt:21, rodriguez:22}. 
However, interpretation of the supposed failed SNe transients and their progenitors is not without controversy and continuing observations are needed to settle the debate \citep{murphy:18, humphreys:19, burke:20, bear:22}.

Historically, SN surveys seek for the sudden appearance of new light sources in the sky. 
However, to discover failed SNe one has to look for the opposite effect: a suddenly ``disappearing'' source. 
Such survey was proposed by \citet{kochanek:08} noticing that within a distance of $10\unit{Mpc}$ there are $\simeq10^6$ red supergiants (RSGs) and that these stars undergo a core-collapse event within $\simeq10^6\unit{yr}$. 
Thus, within a few years observing stars within $10\unit{Mpc}$, a survey would be likely to identify one or even a few stars that end their lives either \textit{with a bang (supernova) or a whimper (fall out of sight)} \citep{kochanek:08}. 
More recently, \citet{tsuna:21} proposed a survey to search for soft x-ray emission from collisions of the relatively slow, $\sim600-800\unit{km\,s}^{-1}$, and low-mass, $\sim0.1\,M_\odot$, ejecta of failed CCSNe from blue supergiant stars (BSG) in the Large Magellanic Cloud (LMC) with the circumstellar medium (CSM). 
Even though the expected rate of BSG failed SNe in the LMC is of order $10^{-4} \unit{year}$, these sources are expected to emit detectable x-rays for $10^3$ to $10^4\unit{years}$, depending on mass loss rate before core collapse for typical BSG ejecta \citep{tsuna:21}. 
The identification of these CCSN progenitors and determination of their outcome would place new constraints on the rate of core-collapse events as well as the branching ratio of failed SNe. 
To date, the failed SN rate has been estimated by \citet{neustadt:21} to be $f=0.16^{+0.23}_{-0.12}$ of all CCSNe and by \citet{byrne:22} to be $f<0.23$ for sources with absolute magnitude $<-14$.

Another reason to identify CCSNe progenitors, especially of failed CCSNe, is to shed light on the ``red supergiant problem''. 
SN surveys by \citet{smartt:09} and \citet{williams:14} have shown that RSGs that are SN progenitors seem to be limited to $M_{\rm ZAMS}\lesssim16.5\,M_\odot$ and $M_{\rm ZAMS}\lesssim20\,M_\odot$, respectively, while some RSGs have $M_{\rm ZAMS}\simeq25\,M_\odot$. 
A possible solution to this problem is that stars in the range $17-20\,M_\odot\lesssim M_{\rm ZAMS}\lesssim25\,M_\odot$ undergo failed CCSN events. 
However, detection selection effects or distinct stellar evolution pathways for stars in the $17-20\,M_\odot\lesssim M_{\rm ZAMS}\lesssim25\,M_\odot$ range could account for the discrepancy \citep{kochanek:08, smartt:09, bear:22}. 
Therefore, unambiguous determination of a failed SN from a RSG progenitor could settle the debate around the RSG problem.

While direct observations are lacking, theoretical and computational works are necessary to evaluate the possible signatures of failed SN and to unequivocally match them to surveys. 
Analytic estimates by \citet{piro:13} show that the response of a RSG to the sudden loss of a few $0.1M_\odot$ in gravitational mass by the PNS is a shock breakout with luminosity $\sim10^{40}-10^{41}\unit{erg\,s}^{-1}$ that lasts $\sim 3-10\unit{days}$. 
\citet{lovegrove:13}, also studying RGS models, demonstrated that failed CCSNe are capable of unbinding the hydrogen envelope of RSGs; leading to faint, red, long-duration observable transients with luminosity $\sim10^{39}\unit{erg\,s}^{-1}$ that lasts approximately a year. 
A main difference between the timescales of the events as predicted by \citet{piro:13} and \citet{lovegrove:13} is that in the former the transient is powered by the kinetic energy of the ejected mass while in the latter the ejected envelope emits its energy via hydrogen recombination. 
Numerical simulations by \citet{lovegrove:17} of the light curve and spectra of failed CCSNe from two RSG progenitors with $M_{\rm ZAMS}=15\,M_\odot$ and $25\,M_\odot$ considering a range of explosion energies show that the peak bolometric luminosity can be in the range $\sim10^{39} - 10^{44}\unit{erg\,s}^{-1}$ depending on explosion energy, which is a function of the equation of state (EOS) and the pre-SN progenitor core compactness, structure, and mass. 
An increase in the maximum mass supported by the PNSs before BH collapse, which is expected if the EOS of dense matter is stiff, results in a net increase of neutrino emission and a stronger shock. 
Also discussed by \citet{lovegrove:17} is the importance of accurately modeling stellar atmosphere, opacity, and ambient medium to determine the observational prospects of failed SN with current and forthcoming missions.

The failed CCSN mass ejection mechanism has also been analyzed analytically by \citet{coughlin:18} using linear perturbation theory. 
Their predictions for the stellar response to the sudden loss of gravitational mass in the core agree well with spherically-symmetric simulation results of \citet{fernandez:18, ivanov:21}, respectively, \citetalias{fernandez:18} and \citetalias{ivanov:21} from now on. 
While \citetalias{fernandez:18} used a parametric prescription to approximate the PNS neutrino mass loss using the non-relativistic \textsc{Flash} code \citep{fryxell:00, dubey:09}, \citetalias{ivanov:21} performed detailed numerical simulations of the PNS evolution up to BH formation using the general-relativistic \textsc{GR1D} code \citep{oconnor:15} and mapped the result to \textsc{Flash} to gauge the ejecta properties. 
Both works showed that mass ejection in failed CCSNe is sensitive to pre-SN progenitor structure, the dense-matter EOS, and, to a smaller degree, to the timescale of gravitational mass loss by neutrino emission, and, might not occur at all for some progenitors.

In this work we perform simplified numerical simulations of failed CCSNe. 
First, we develop a model to estimate neutrino emission and BH formation times for CCSNe based on spherically-symmetric simulations using \textsc{Flash} \citep{fryxell:00, dubey:09, couch:13, oconnor:18} and M1 neutrino transport \citep{oconnor:15}. 
Extending the simulations of \citet{schneider:20} (\citetalias{schneider:20} henceforth) to a few other progenitors, we show that (1) the evolution of the neutrino luminosity can be well fitted by a function of the accretion rate onto the PNS and its baryonic mass and that (2) the entropy inside the PNS, which determines BH formation for an EOS, is well approximated by a function of the mass of the PNS and the time since the PNS has formed. 
Together, these two approximations allow us to simulate failed CCSNe by replacing the PNS and BH physics by a simple model that includes accretion of the supersonically infalling outer layers of the star and the loss of gravitational mass due to neutrino emission without having to solve computationally expensive neutrino transport equations.

Employing our model for PNS neutrino emission and BH formation we perform CCSNe simulations in spherical symmetry for a wide range of pre-SN progenitor stars found in the literature. 
We watch as a sound pulse forms in the outer layers of the star due to the loss in core gravitational mass and run our simulations well past BH formation until the sound pulse either becomes a shock and unbinds from the star or falls back into the BH. 
By using a large set of simulations we determine the dependence of the mass ejecta and its energy as a function of the pre-SN progenitor properties and EOS.

We discuss our model of neutrino emission and PNS evolution during CCSNe in Section~\ref{sec:model}. 
To validate our model, in Section~\ref{sec:vs} we compare PNS evolution until BH formation using our template to results obtained with significantly more computationally expensive core-collapse simulations that use M1 neutrino transport. 
In Section~\ref{sec:results} we use our parameterized model to predict the impact of progenitor structure and EOS on the properties of failed CCSNe. 
We conclude in Section~\ref{sec:conclusions}.

\section{Model}
\label{sec:model}

We first review the main result of \citetalias{schneider:20} and, then, use the insights gained to parameterize the PNS entropy evolution and its neutrino emission.

\subsection{Black hole formation}
\label{ssec:BH formation}

In \citetalias{schneider:20} we have shown that for a given pre-SN progenitor the PNS gravitational mass and its \textit{most common entropy}\footnote{The \textit{most common entropy} $\tilde{s}$ is defined as the value of the approximately flat PNS entropy that contains the largest amount of mass.}, simply referred to as entropy from now on, evolve almost independently of the EOS up to the point of BH formation. 
However, the BH formation time and the EOS of dense matter. 
The latter was hinted at by \cite{hempel:12, steiner:13} discussing the core collapse of a single progenitor star and many EOSs.

\begin{figure}[htb]
\includegraphics[width=0.47\textwidth]{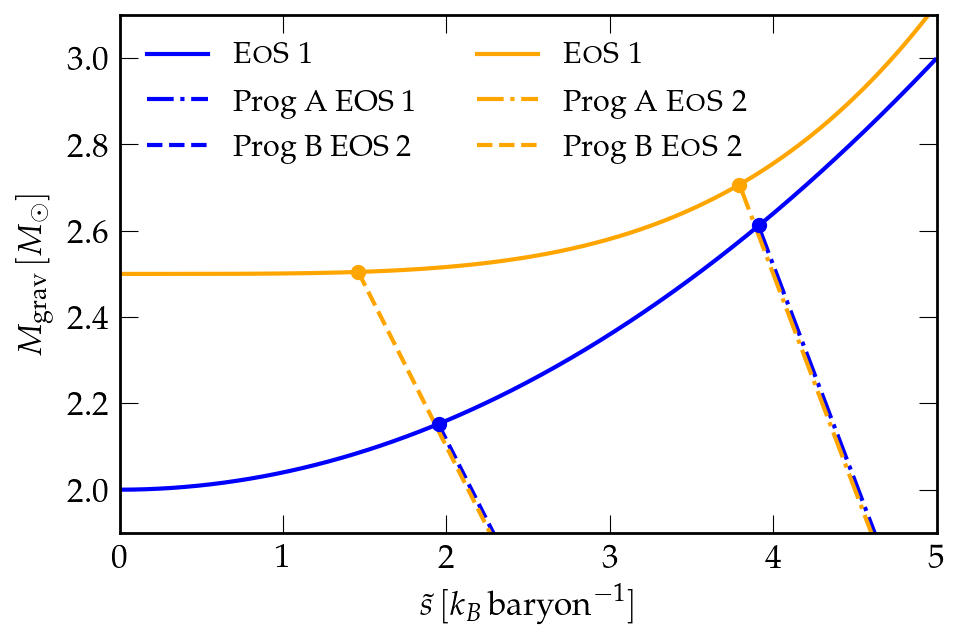}
\caption{\label{fig:scheme} PNS mass-entropy evolution scheme up to the point of BH formation for two different EOSs and two pre-SN progenitor stars. 
Tracks for high-compactness progenitor A (dash-dotted lines) and low-compactness progenitor B (dashed lines) start at low $M_\rgrav$ and move to higher mass and lower entropy as matter is accreted and neutrinos are emitted. 
The PNS collapses into a BH earlier (lower PNS mass) for the softer \textsc{EOS 1} than the stiffer \textsc{EOS 2}.
}
\end{figure}

For clarity and simplicity we show in Figure~\ref{fig:scheme} a scheme with four PNS mass-entropy evolution tracks for the core collapse of two different pre-SN progenitor stars, each one evolved considering two different dense-matter EOSs. 
In the scheme, dashed and dot-dashed tracks show the entropy evolution of the PNS as a function of the PNS gravitational mass up to the point of BH formation. 
The latter depends on the maximum mass supported by a PNS at finite entropy, shown as solid lines in the plot. 
In the plot, progenitor A is a massive star with a compactness larger than that of progenitor B. 
Hence, as the iron core of progenitor A collapses it forms a PNS that accretes matter faster than the PNS formed by the core collapse of progenitor B. 
Faster accretion leads to higher energy deposition onto the PNS and, thus, higher temperatures and entropies inside the PNS. 
This explains why progenitor A has a higher entropy than progenitor B for the same PNS mass. 
Nevertheless, the entropy of both PNSs decrease as they become more massive, as entropy is carried away by neutrino emission. 
In the case of pre-SN progenitors with mass-entropy tracks similar to those of progenitor A, due to fast accretion of matter, not much time elapses before the mass of the PNS overcomes the maximum mass that can be supported by the PNS at a given entropy and a BH forms. 
For entropies of the magnitudes shown for progenitor A, the time from core bounce to BH formation is $t_{\rm BH}-t_{\rm bounce}\lesssim 0.5\unit{s}$. 
Meanwhile, PNSs formed in CCSNe that have similar $M_\rgrav-\tilde{s}$ tracks to progenitor B accrete matter slower due to the lower accretion rates onto the PNS and $t_{\rm BH}-t_{\rm bounce} \gtrsim 1\unit{s}$.

As shown in Figure~\ref{fig:scheme}, the PNS mass-entropy evolution for a given progenitor  does not depend significantly on the EOS up to the point of BH formation, shown by filled circles in the scheme. 
The moment the BH forms, however, is EOS dependent. 
In Figure~\ref{fig:scheme} we compare two different EOSs: 
\textsc{EoS 1} is a softer EOS than \textsc{EoS 2} at lower temperatures (low entropy), while both EOSs have similar stiffness at higher temperatures (high entropy). 
Therefore, for the more (less) compact progenitor A (B) the difference in PNS gravitational mass at the moment of BH formation is relatively small (large), $\simeq 0.1\,M_\odot$ ($\simeq 0.4\,M_\odot$). 
Although there are only a few constraints on the maximum mass supported by a PNS at a given entropy, most models predict a maximum supported mass that increases with entropy \citep{schneider:20, steiner:13}.

The picture outlined above highlights that it would be difficult to extrapolate zero temperature NS properties from the observation of the formation of a BH from a CCSN. 
Nevertheless, this picture is useful to simplify the study of failed CCSNe since we are able to create a template for neutrino luminosity and PNS entropy evolution from the outcome of only a few simulations. 
With a reasonable template we are able to determine time dependent neutrino emission from a PNS as well as the BH formation time for different progenitors and EOSs without having to to solve computationally expensive neutrino transport equations. 
We discuss this now.

\subsection{Protoneutron star entropy model}
\label{ssec:entropy}

To create a simple template for PNS entropy evolution we perform spherically symmetric simulations of CCSNe using the \textsc{Flash}-code \citep{fryxell:00, dubey:09, couch:13, oconnor:18}. 
Our simulations employ the SRO baseline EOS of \citet{schneider:19a, eggenbergerandersen:21} and the neutrino transport library \textsc{NuLib} of \citet{oconnor:15} as discussed in detail in \citetalias[Section 2.3]{schneider:20}.

\begin{table}[hbt]
\centering
\caption{ \label{tab:presn} Pre-SN progenitor name, zero age main sequence mass $\left(M_{\mathrm{ZAMS}}\right)$, total pre-SN mass at the start of core collapse $\left(M_{\mathrm{cc}}\right)$, iron core mass $\left(M_{\mathrm{Fe}}\right)$, core-compactness parameters $\xi_{2.5}$ \citep{oconnor:11}, envelope compactness $\xi_{\rm env}$, and metallicity $z$ relative to that of the Sun $z_\odot$, $[z]=\log_{10}(z/z_\odot)$ where ``x'' means no metals. 
Compactness, $\xi_{2.5}$, values are for pre-SN progenitors at the start of collapse and, thus, differ from the ones of \citet{oconnor:11}, which were computed for a single EOS at the moment of core bounce. 
}
\begin{tabular}{lDDDDDc}
\hline
\hline
{Name} & 
\multicolumn2c{$M_{\mathrm{ZAMS}}$} &
\multicolumn2c{$M_{\mathrm{cc}}$} &
\multicolumn2c{$M_{\mathrm{Fe}}$} &
\multicolumn2c{$\xi_{2.5}$} & 
\multicolumn2c{$\xi_{\rm env}$} & 
\multicolumn1c{$[z]$} \\
{ } & 
\multicolumn2c{$[M_\odot]$} & 
\multicolumn2c{$[M_\odot]$} &
\multicolumn2c{$[M_\odot]$} &
\multicolumn2c{ } & 
\multicolumn2c{ } & 
\multicolumn1c{} \\
\hline
\multicolumn7l{\citet{fernandez:18}\footnote{The  MESA \citep{paxton:11, paxton:13, paxton:15, paxton:18, paxton:19} inlists to generate the CCSN progenitors were obtained from \url{https://bitbucket.org/rafernan/bhsn_mesa_progenitors/src/master/}.}} \\
\hline
\decimals
\texttt{R12z00}    & 12.0 & 10.0 & 1.45 & 0.151 &  0.009 &  0. \\
\texttt{W26z00}    & 26.0 & 11.9 & 1.59 & 0.214 & 10.8   &  0. \\
\texttt{Y25z-2}    & 25.0 & 23.0 & 1.62 & 0.256 &  0.024 & -2. \\
\texttt{W40z00}    & 40.0 & 10.3 & 1.78 & 0.370 & 27.1   &  0. \\
\texttt{R15z00}    & 15.0 & 10.8 & 1.50 & 0.239 &  0.010 &  0. \\
\texttt{B25z00}    & 25.0 & 11.7 & 1.60 & 0.336 &  0.12  &  0. \\
\texttt{B30z-2}    & 30.0 & 16.0 & 1.72 & 0.337 &  0.11  & -2. \\
\texttt{Y22z00}    & 22.0 & 11.1 & 1.85 & 0.546 &  0.016 &  0. \\
\texttt{W50z00}    & 50.0 &  9.2 & 1.91 & 0.554 & 21.9   &  0. \\
\texttt{B80z-2}    & 80.0 & 55.2 & 3.32 & 0.972 &  0.79  & -2. \\
\hline
\multicolumn7l{\citet{woosley:07}} \\
\hline
\texttt{s50}   & 50.0 &  9.76 & 1.50 & 0.216 & 30.1   & 0. \\
\texttt{s25}   & 25.0 & 15.8  & 1.60 & 0.326 &  0.011 & 0. \\
\texttt{s40}   & 40.0 & 15.3  & 1.83 & 0.534 &  1.40  & 0. \\
\hline
\multicolumn7l{\citet{woosley:02}}  \\
\hline
\texttt{z11}  & 11.0 & 11.0 & 1.25 & 0.005 & 0.591 & x \\
\texttt{z12}  & 12.0 & 12.0 & 1.36 & 0.011 & 0.997 & x \\
\texttt{z13}  & 13.0 & 13.0 & 1.45 & 0.025 & 1.255 & x \\
\texttt{z14}  & 14.0 & 14.0 & 1.40 & 0.041 & 1.354 & x \\
\texttt{z25}  & 25.0 & 25.0 & 1.81 & 0.385 & 2.148 & x \\
\texttt{u40}  & 40.0 & 40.0 & 1.90 & 0.633 & 0.457 & -4 \\
\texttt{u75}  & 75.0 & 74.1 & 2.03 & 0.873 & 0.299 & -4 \\
\hline
\multicolumn7l{\citet{limongi:06} (LC06A)} \\
\hline
\texttt{s60}  &  60.0 & 16.9 & 1.63 & 0.424 & 0.081 & 0. \\
\texttt{s80}  &  80.0 & 22.4 & 1.67 & 0.481 & 0.111 & 0. \\
\texttt{s120} & 120.0 & 30.5 & 1.91 & 0.534 & 0.143 & 0. \\
\hline
\multicolumn7l{\citet{sukhbold:16}} & \multicolumn2l{$[10^{-4}]$}  \\
\hline
\texttt{s9.0}  &  9.0 & 8.75 & 1.32 & 0.382 & 0.021  & 0. \\
\texttt{s10.0} & 10.0 & 9.68 & 1.34 & 1.99  & 0.019  & 0. \\
\hline
\hline
\end{tabular}
\end{table}

We simulate the core collapse of many pre-SN progenitors with core-compactness $\xi_{2.5}$ spanning the range $\sim10^{-4.5}$ to $\sim1$ and envelope compactness $\xi_{\rm env}$ in the range $\sim0.01$ to $\sim30$. 
Core-compactness is defined as in \citet{oconnor:11}, 
\begin{equation}\label{eq:xi_M}
 \xi_M = \left.\frac{M/M_\odot}{R(M_{\rbaryon}=M)/1\,000\unit{km}}\right|_{t=t_{\rm cc}}\,,
\end{equation}
and envelope compactness as in \citetalias{fernandez:18},
\begin{equation}\label{eq:xi_env}
 \xi_{\rm env} = \left.\frac{M_{\rm cc}/M_\odot}{R_{\rm cc}/R_\odot}\right|_{t=t_{\rm cc}}\,,
\end{equation}
where $M_{\rm cc}$, $R_{\rm cc}$, and $t_{\rm cc}$ are the progenitor mass, radius, and time at the start of core collapse. 
The list of progenitors used to create a PNS mass-entropy evolution template and their properties are shown in Table~\ref{tab:presn}. 
For each simulation we compute the entropy $\tilde{s}$ inside the PNS for 100 equally separated time instants from core bounce until a BH forms or, in the case of low-compactness models, until $t_{\rm{simulation}}-t_{\rm cc}=10\,\unit{s}$. 
We then fit the entropy as a function of two variables using a second-order-splines interpolation scheme \citep{virtanen:20}.
The variables are 
(1) base-10 logarithm of the simulation time after core bounce, $\log_{10}t$ where $t=t_{\rm{simulation}}-t_{\rm{bounce}}$, and 
(2) the baryonic mass inside a radius of $500\unit{km}$, $M_{500}$. 
Here $t_{\rm{simulation}}$ is the time from the start of the simulation and we approximate the bounce time $t_{\rm{bounce}}$ as the time after the central density of the star reaches $10^{12}\unit{g\,cm}^{-3}$.

\begin{figure}[htb]
\includegraphics[width=0.50\textwidth]{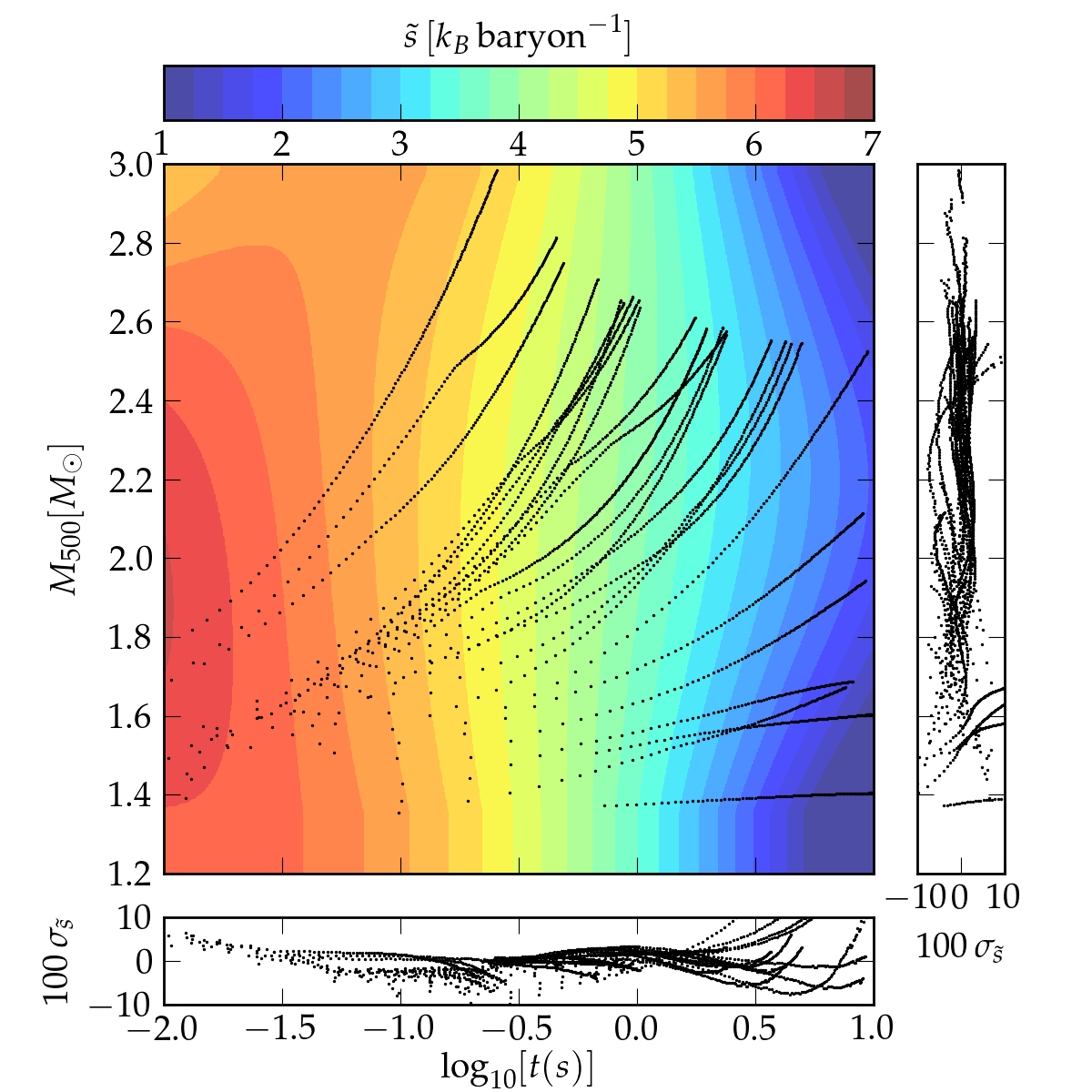}
\caption{\label{fig:entropy} Entropy $\tilde{s}$ fit based on CCSN simulations of the pre-SN progenitors described in Table~\ref{tab:presn}. 
The values of $t$ and $M_{500}$ used to fit the entropy are shown as black dots in the main plot. 
The deviations $\sigma_{\tilde s}$ between the fit and the entropy computed from the simulations are shown in the bottom and right plots and are usually 5\% or better. 
}
\end{figure}

In Figure~\ref{fig:entropy}, we show the results of the entropy fit $\tilde{s}(t,M_{500})$, the tracks in $t-M_{500}$ space for each core-collapse simulation using the progenitors of Table~\ref{tab:presn}, and the deviations between the fit and the simulation results. 
First, we observe that in the first second after bounce, the PNS entropy is, to first order, only a function of time after core bounce, \ie almost independent of progenitor compactness or PNS mass. 
Thus, for compact progenitors that form BHs in $t\lesssim1\unit{s}$, it would have been sufficient to fit the entropy as a function of time. 
However, since the trend breaks down for PNSs that take longer than 1\unit{s} to collapse into a BH, we opt for a more robust two variable fit. 
Note that the fit for massive PNSs when $M_{500}\gtrsim2.5\,M_\odot$ and $t\gtrsim1\unit{s}$ predicts decreasing entropy with increasing mass. 
This may be unrealistic and simply an artifact of the fitting as these regions in phase space are not populated by our simulations. 
We also notice that the deviation in entropy between the fit and the simulations are often within 5\% of each other, with the largest deviations taking place near the end of each. 
Furthermore, changes of $\sim10\%$ in the estimated PNS entropy do not alter significantly the BH formation time and its initial mass. 
For very-low compactness progenitors, which take significantly longer than 10\unit{s} to form a BH, we expect changes in entropy to cause an even smaller relative error in the BH formation time, as the maximum mass supported by the EOS does not change more than $0.1\,M_\odot$ compared to that of a cold NS. 
In Section~\ref{ssec:simulations} we discuss how we approach neutrino emission for supernova progenitors that take more than 10\unit{s} after core bounce to form a BH.

\subsection{Neutrino emission model}
\label{ssec:neutrino}

Similar to what is done for the PNS entropy, we also fit the neutrino luminosity $L_{\nu}$ for the three neutrino species considered in our simulations, $\nu = \nu_e, \,\bar\nu_e, \, \nu_x$ where $\nu_x$ includes contributions from $x=\mu$, $\bar\mu$, $\tau$, and $\bar\tau$ neutrino species. 
The neutrino luminosity fits are done for each neutrino species independently and are interpolated as a function of (1) the mass inside a radius of 500\,km from the center of the star, $M_{500}$, and (2) the accretion rate at 500\,km, $\dot{M}_{500}$, using a radial-basis function interpolation scheme. 
We chose the variables $M_{500}$ and $\dot{M}_{500}$, instead of $\log_{10}t$ and $M_{500}$ as in the entropy template, because they provide a considerable better fit to the neutrino luminosity and are similar to the variables used in the analytic neutrino luminosity approximation of \citet{lovegrove:13} and \citetalias{fernandez:18}.

\begin{figure}[htb]
\includegraphics[width=0.50\textwidth]{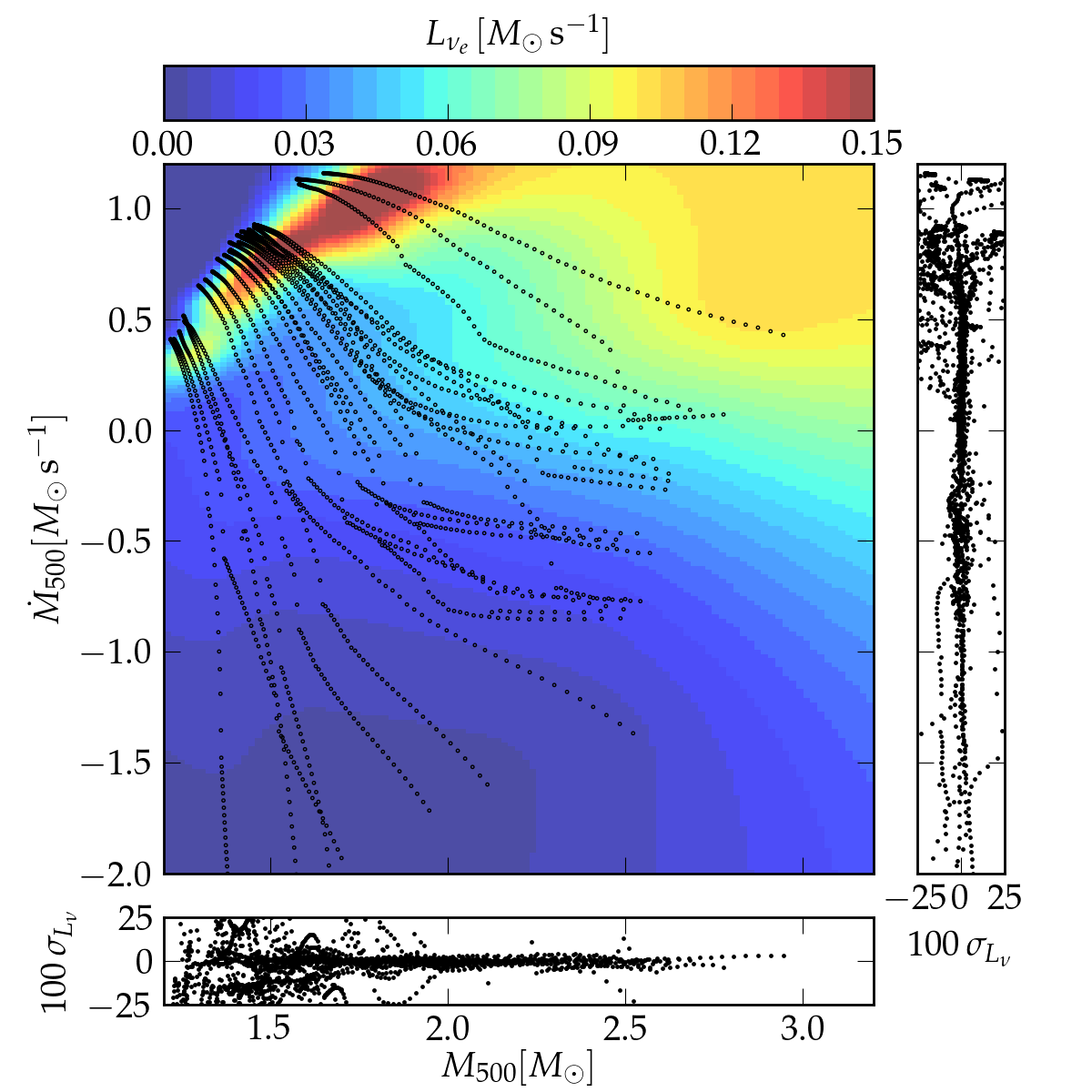}
\caption{\label{fig:Lnue} Electron neutrino luminosity $L_{\nu_e}$ fit based on simulations of pre-SN progenitors described in Table~\ref{tab:presn}. The values of $M_{500}$ and $\dot{M}_{500}$ used to fit the neutrino luminosity are shown as black dots in the main plot. 
The deviations $\sigma_{L_{\nu_e}}$ between the fit and the luminosity computed from the simulations are shown in the bottom and right plots. 
Deviations are usually $\lesssim15\%$, except for regions in the parameter space where accretion is high or drops very quickly due to accretion of a shell boundary. 
Note that $0.1\,M_\odot\simeq1.8\times10^{53}\unit{erg}$.}
\end{figure}

In Figure~\ref{fig:Lnue}, we show the electron neutrino luminosity fit, $L_{\nu_e}$, and the path traced in $M_{500}-\dot{M}_{500}$ by the PNSs generated by the collapse of the progenitors described in Table~\ref{tab:presn}. 
The interpolated luminosity is usually within 20\% of the simulated values although large deviations do occur at selected times. 
Specifically, during the neutronization burst, when the accretion rate is $\mathcal{O}(10\,M_\odot\unit{s}^{-1})$, and at times where the accretion rate has a sharp decrease due to the accretion of a shell boundary at 500\unit{km}. 
Since such moments are quite brief with respect to the BH formation time, their contributions to the total neutrino luminosity are small. 
We observe similar trends for the luminosity fits of the other two neutrino species, $\bar\nu_e$ and $\nu_x$, except for the lack of a neutronization burst.

Our method to predict neutrino luminosity during CCNSe has other limitations. 
First, we observe that neutrino luminosity predicted using our simplified model can, in some instances, deviate significantly from the full M1 neutrino-transport scheme calculations. 
This effect is stronger for the heavy neutrino luminosity $L_{\nu_x}$ and is especially true for stiff EOSs, which increase the time to form BHs. 
This is in part because to simulate core collapse with stiffer EOSs, where BH formation occurs for larger PNS masses, we have to extrapolate the neutrino luminosity fits into regions of the $M_{500}-\dot{M}_{500}$ parameter space that our template, based on results obtained with the SRO baseline EOS, does not cover. 
We discuss these effects further in Section~\ref{sec:vs}.

Low-compactness progenitors accrete matter very slowly and, therefore, take a long time to collapse into a BH, $t\gtrsim10\unit{s}$ after core-bounce. 
Since our model is limited to fits where $\dot{M}_{500}>0.01\,M_\odot\unit{s}^{-1}$, extrapolation from this region may lead to relatively high neutrino luminosity at late times. 
In fact, while accretion leads to an increase in baryonic mass, extrapolating from the parameter space probed can result in unrealistically high or low neutrino luminosity, which can lead to BHs forming too fast (no neutrino emission) or never forming at all (neutrino luminosity larger than accretion rate). 
To prevent this unrealistic scenarios, we cap the neutrino luminosity $L_{\nu}$ for each species to $\dot{M}_{500}/10$. 
This scenario only affects low compactness progenitors several seconds after core bounce, when neutrino emission is expected to dissipate any thermal heat gain due to accretion in time scales much shorter than the accretion timescales. 
Thus, to a good approximation, we may consider such NSs to be cold, zero temperature, and slowly increasing their mass. 
We remark that for most EOSs and cold NSs with baryonic mass $M_{\rm{baryon}}\gtrsim1.5\,M_\odot$, $dE_{\rm{bind}}/dM_{\rm{baryon}}\sim0.30-0.35$, where 
$E_{\rm{bind}} = M_{\rm{baryon}} - M_{\rm{grav}}$ is the binding energy of the cold NS. 
Thus, by capping the neutrino emission rates to a combined luminosity of $\sim0.3\,\dot{M}_{500}$ we obtain a decent estimate of neutrino emission at late times.

\subsection{Simulations}
\label{ssec:simulations}

The parameterizations discussed above allow us to simulate the outcome of failed CCSNe without having to solve the computationally expensive neutrino transport equations.  
This, we are able to simulate failed SN from the start of core collapse until shock breakout, if it occurs, which may be minutes to years after BH forms depending on the progenitor star. 
The simulations are performed using the \textsc{Flash4} code \citep{fryxell:00, dubey:09,couch:13, oconnor:18} in spherical symmetry. 
The code is adapted to estimate the PNS entropy $\tilde{s}$ and neutrino luminosity $L_\nu$ using the parameterized templates described above. 
Our runs are divided in three stages discussed below. 
In every run, we use a spherical grid with adaptive mesh refinement that extends out to $2\times10^{12}\unit{cm}$ in the first two stages and then mapped onto a grid that extends to $2.048\times10^{15}\unit{cm}$ in the last stage. 
In all stages, on the coarsest level, there are 256 grid zones and we allow up to 20 total levels of refinement resulting in a smallest
grid zone with a length of 149\unit{m} for the first two stages and 153\unit{km} for the last stage.

\subsubsection{Stage 1: Before core bounce}
\label{sssec:bounce}

We evolve the core-collapse of the pre-SN progenitor until the central density reaches $\rho_c=10^{12}\unit{g\,cm}^{-3}$. We do not use any neutrino transport and set $t=0$ (core bounce) at the end of this stage. 
Simulating this short period before core bounce is necessary to accurately reproduce the mass accretion rate onto the PNS after core bounce, which is crucial to use the neutrino luminosity template discussed in Section~\ref{ssec:neutrino}.

\subsubsection{Stage 2: From core bounce to BH formation}
\label{sssec:BH}

At core bounce we replace the inner $r=r_0=100\unit{km}$ of the simulation volume by a ``hole'' with constant density $\rho_\rhole$. 
The density $\rho_\rhole$ is computed assuming that the hole contains a mass that is equal to half of the mass between $r$ and $2r$, guaranteeing that $\rho_\rhole\ll\rho(r_0)$ for $r=r_0$ and, thus, the mass just outside the hole free-falls onto the PNS. 
This reproduces the expected accretion rate for CCSNe that do not lead to successful explosion, as is the case for all spherically-symmetric simulations for the pre SN progenitors probed in this work\footnote{This approach ignores that a shock-wave is created at the surface of the PNS and travels outwards before stalling and falling back onto the PNS. 
However, this is not relevant as this region is sonically disconnected from the outer layers by the supersonic accretion flow, where the sound pulse we are interested in forms.}. 
The ``missing'' gravitational mass from decreasing the density in the region within the hole radius $r$ is placed as a point mass at the origin. 
In order to increase the time step of the simulation as the system evolves, the hole size is increased with a constant speed such that $r=r_0+vt$, where we set $v=10\unit{km\,s}^{-1}$. 
The point mass at the origin and the hole density are evolved according to the hole size evolution and the neutrino emission computed from our templates.

We estimate the gravitational mass inside the 500\unit{km} sphere used in our templates to be $M_{\rgrav}(t) = M_{500}(t)-E_\nu(t)$, where $E_\nu(t)=\sum_\nu\int_0^t L_\nu(t') dt'$ is the total energy emitted in neutrinos and the sum runs over the three neutrino species considered. 
This approach is justified as, to a very good approximation, neutrinos are the only source of gravitational mass loss. 
By tracking the entropy and gravitational mass evolution we determine the moment a BH forms from $M_{\rgrav}(t)=M_{\rgrav}^{\rmax}(s(t))$ where $s(t)$ is the template estimate for the PNS entropy.  
The point of BH formation depends on $M_{\rgrav}^{\rmax}(s)$, which can be set to some fixed value or be computed for a desired EOS, see Figure~\ref{fig:scheme}. 
Low-compactness progenitors may take significantly longer than 40\unit{s} to collapse into a BH, the time when the hole radius reaches the 500\unit{km} used in our templates. 
When this happens, we simply compute the baryonic mass and accretion rate at the hole radius $r$ instead of $500\unit{km}$ as the time for matter accreted at $r$ to reach $500\unit{km}$ is significantly shorter than the dynamical time scale of the system.

\subsubsection{Stage 3: after BH formation}
\label{sssec:late}

Once a BH forms, the neutrino luminosity is set to zero and the total gravitational mass loss from neutrino emission is $\delta M_\rgrav=E_\nu(t_\rBH)$. 
Because of the change in gravitational mass of the inner core, hydrodynamic equilibrium in the outer layers of the pre-SN progenitor is disturbed \citep{nadyozhin:80, lovegrove:13, coughlin:18}. 
This perturbation creates a pressure wave that propagates outwards towards the surface of the star and may result in some mass ejection, even in the case of failed SNe. 
To determine the mass ejection and its energy, we follow the evolution of the system until the pulse turns into a shock and leaves the star or the pulse velocity becomes negative, indicating the pulse will fall back into the BH. 
We limit the hole radius to $r_\rmax=2\times10^{7}\unit{km}$, at which point we fix $r=r_\rmax$. 
As in \citetalias{fernandez:18, ivanov:21} we (1) fill the region outside the star with a constant density ambient medium in hydrostatic equilibrium for numerical reasons and (2) map pressure, density, and proton fractions from the progenitor to recover the remaining thermodynamic variables using the Helmholtz EOS \citep{timmes:00} to minimize transients. 
The ambient medium is set to a hydrogen gas with density $\rho_\ramb=\{10^{-18}, 10^{-16}, 10^{-14}, 5\times10^{-13}\}\unit{g\,cm}^{-3}$ for stars with radii $R > \{10^{8}, 10^{7}, 10^{6}, 10^{5}\}\unit{km}$, respectively. 
To achieve densities below $10^{-10}\unit{g\,cm}^{-3}$, the Helmholtz EOS as implemented in \textsc{Flash} is extended to $10^{-20}\unit{g\,cm}^{-3}$ using a version of the publicly available \citet{timmes:99} EOS code\footnote{To handle the accuracy needed at very low densities we extend the \citet{timmes:99} EOS code, {\tt timmes.tbz} found in \url{https://cococubed.com/code_pages/eos.shtml}, to quadruple precision.}. 
Similar to \citetalias{fernandez:18,ivanov:21}, simulations are terminated once the pulse reaches the atmosphere surrounding the star and its temperature drops to within 1\% of the EOS table lower limit, $T_{\rm low}=10^4\unit{K}$ or if the pulse leaves the simulation volume. 
The spatial resolution of our simulations in the outer regions of the star, $r>10^8\unit{cm}$, are $\Delta r/r=4\times10^{-3}$, similar to the low resolution runs of \citetalias{ivanov:21}. 
We did not perform detailed resolution studies since \citetalias{ivanov:21} showed that resolution is a smaller source of uncertainty than changes in the EOS.

\section{Template accuracy}
\label{sec:vs}

Using the methods described in Section~\ref{sec:model}, we assess the accuracy of our model in describing the core collapse of the $40\,M_\odot$ solar metallicity pre-SN progenitor of \citet{woosley:07}, \texttt{s40WH07}, using a variety of EOSs. 
Besides the baseline EOS of \citetalias{schneider:20}, used to construct our PNS entropy and neutrino emission templates, we explore 21 other EOSs found in the literature\footnote{The stiff NL3 \citep{hempel:12} and LS$_{375}$ \citep{lattimer:91, oconnor:11} EOSs were not considered as they would require us to rely on our neutrino emission template in regions of $M_{500}-\dot{M}_{500}$ parameter space where it becomes unreliable.}. 
The full set of EOSs contains the baseline SRO EOS, SRO$_{0.75}$ which has $m^\star/m_n=0.75$, and its stiff variant SRO$_{0.55}$ with $m^\star/m_n=0.55$ and soft variant SRO$_{0.95}$ with $m^\star/m_n=0.95$ \citep{schneider:19, eggenbergerandersen:21}. We also include the APR, APR$_{\rm{LDP}}$, NRAPR, and SkAPR EOSs \citep{akmal:97, akmal:98, steiner:05, schneider:19a}; two variants of the Lattimer and Swesty (LS) EOS with incompressibilities $K_\rsat = 180$ (LS180) and $220\unit{MeV\,baryon}^{-1}$ (LS220) \citep{lattimer:91, oconnor:11}; the DD2, FSU-Gold, TM1, and TMA EOSs \citep{hempel:12}; the IU-FSU EOS \citep{fattoyev:10}; the two DD2 variants that include hyperons: BHB$_{\Lambda}$ and BHB$_{\Lambda\phi}$ \citep{banik:14}; the H.~Shen EOS \citep{shen:98a} and its variant including $\Lambda$ hyperons \citep{shen:11}; the SFHo and SFHx EOSs \citep{steiner:13}; the Togashi EOS \citep{togashi:17}; and the Furusawa EOS \citep{furusawa:17}.

\begin{figure}[htb]
\includegraphics[width=0.47\textwidth]{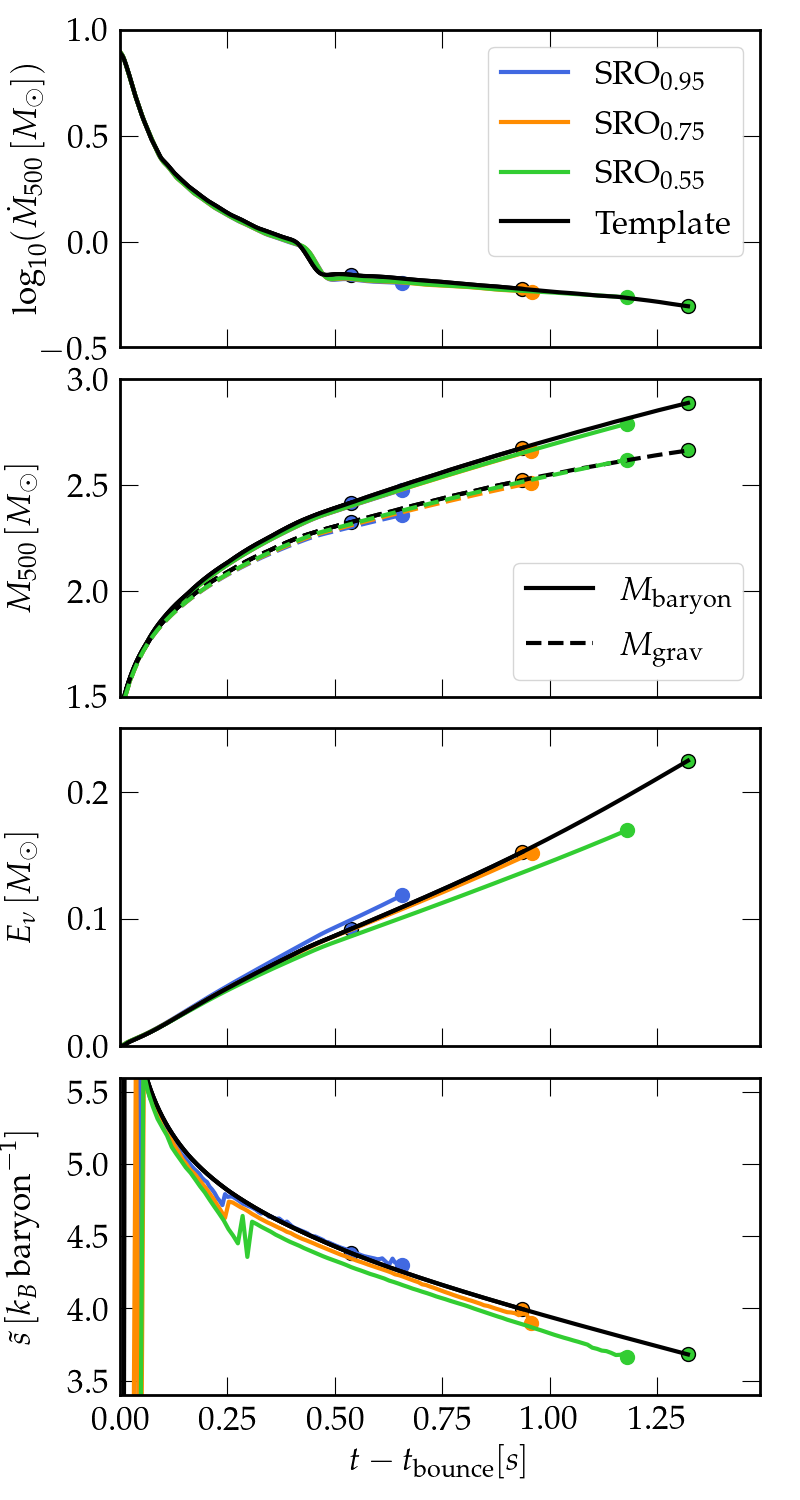}
\caption{\label{fig:s40_SRO} Comparisons of time evolution of mass accretion rate at $500\unit{km}$ $\dot{M}_{500}$ (first panel), baryonic $M_{\rbaryon}$ and gravitational $M_{\rgrav}$ masses inside 500\unit{km} (second panel), the energy emitted in neutrinos $E_{\nu}$ (third panel) and entropy $\tilde{s}$ (fourth panel) obtained using M1 neutrino transport and the parameterizations discussed in Section~\ref{sec:model}. Results are for the SRO EOSs with $m^\star/m_n=$0.55, 0.75, and 0.95 \citep{schneider:19a}. Simulation endpoints are marked by circles. Deviations in final masses and entropy are at most a few percent, while deviations in total mass loss by neutrino emission can be up to $\sim20\%$.}
\end{figure}

In Figure~\ref{fig:s40_SRO}, we compare results of simulations that use the simplified templates, discussed in Section~\ref{sec:model}, to simulations performed using a full M1 neutrino-transport scheme for the three SRO EOS variants. 
We show that simulations using the simplified neutrino transport scheme predicted the evolution of the PNS entropy, the integrated neutrino luminosity $E_{\nu}$, and the PNS gravitational mass $M_{\rm{grav}}$ within a few percent of their equivalent M1 neutrino-transport runs. 
However, since our template was built for a single EOS the end of the runs occur earlier (later) by up to 20\% of the run time for EOSs softer (stiffer) than the baseline SRO$_{0.75}$ EOS. 
This occurs because the stiffness of the EOS at a given entropy affects the neutrino luminosity: softer EOSs lead to faster PNS contraction and, therefore, faster heating and higher neutrino emission rates \citep{schneider:19a, yasin:20}. 
This effect, though, is not taken into account on our template, which is based solely on $M_{500}$ and $\dot{M}_{500}$. 
Thus, the integrated neutrino luminosity by BH formation time predicted by our template can also differ by up to $\sim20\%$ from the M1 neutrino-transport runs depending on progenitor and EOS used. 
We stress that this $\sim20\%$ difference in BH formation time and total gravitational mass loss (integrated neutrino luminosity) is only a \textit{secondary} source of uncertainty as the total gravitational mass loss due to neutrino emission by BH formation time can differ by a factor of $\sim2$ for different EOSs for the same progenitor \citepalias{schneider:20, ivanov:21}.

\begin{figure}[htb]
\includegraphics[width=0.47\textwidth]{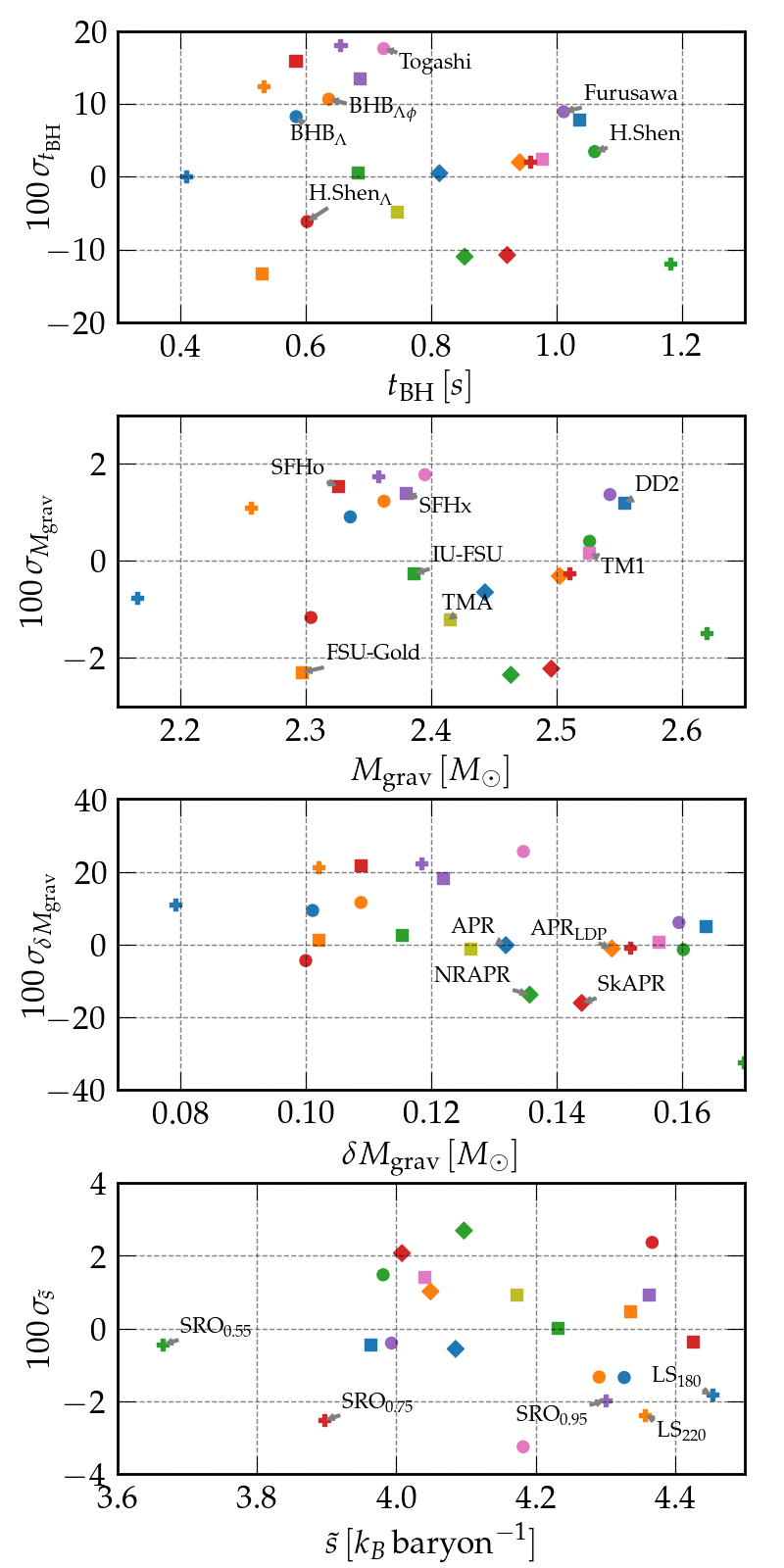}
\caption{\label{fig:s40_sigma} Deviations $\sigma$ at BH formation time between core-collapse simulations for the s40WH07 progenitor of \citet{woosley:07} using M1 neutrino transport and the paramaterized model discussed in Section~\ref{sec:model}. 
We plot deviations in BH formation time $t_\rBH$ (first panel), gravitational mass inside 500\unit{km} $M_{\rgrav}$ (second panel), total PNS gravitational mass loss $\delta M_{\rgrav}$ (third panel), and PNS entropy $\tilde{s}$ (fourth panel), all at $t_\rBH$. 
Plots have the full transport results in the horizontal axis and include deviations for 22 EOSs found in the literature.}
\end{figure}

We extend the analysis above to the other 19 EOSs considered by comparing deviations in observables $A$, \mbox{$\sigma_A = 1 - {A^{\rm (template)}}/{A^{\rm (M1)}}$}, obtained from M1-transport simulations and their counterparts that use our parameterized model. 
In Figure~\ref{fig:s40_sigma}, we plot the deviations between the two approaches for the BH formation time $t_{\rBH}$, gravitational mass $M_{\rgrav}$, PNS gravitational mass $\delta M_\rgrav$ (which is equal to the total integrated neutrino luminosity $E_{\nu}$), and PNS entropy $\tilde{s}$. 
The last 3 quantities are computed $\sim1\unit{ms}$ before black hole formation time $t_\rBH$. 
Although there are EOS-dependent variations in the predictions made for all of these quantities, the deviations are usually only a few percent for $\sigma_{M_\rgrav}$ and $\sigma_{\tilde s}$. Meanwhile, BH formation time and total neutrino luminosity have relative deviations that can be up to a factor of $\sim10$ larger, $\sigma_{t_\rBH}\lesssim20\%$ ($\lesssim100\unit{ms}$) and $\sigma_{E_{\delta M_\rgrav}}\lesssim25\%$.

Moreover, we note that the average neutrino emission rate, $\delta M_\rgrav/t_\rBH$, is also satisfactorily approximated by our template, even if the total emitted neutrino energy by BH formation time deviates by up to 20\% in some cases due to difference in predicted $t_{\rBH}$. 
In fact, except for EOSs that are very stiff in their temperature dependence (SRO$_{0.55}$ EOS) or very soft (SRO$_{0.95}$, FSU-GOLD, LS$_{180}$, LS$_{220}$, SFHo, and Togashi EOSs), the average neutrino emission rate computed from our template is within 5\% of the one obtained from M1-transport simulations. 
Even EOSs that are very soft or very stiff and, thus, have a larger deviation in the predicted PNS binding energy, $\vert\sigma_{\delta M_{\rgrav}}\vert\sim20\%$, have relative deviations in average neutrino emission rate a factor of 2 to 4 lower, $\vert\delta M_\rgrav\vert/t_\rBH\simeq5-10\%$. See for example the slopes of $E_{\nu}$ for the SRO$_{0.55}$ and SRO$_{0.95}$ EOSs in Figure~\ref{fig:s40_SRO}.

From the reasons stated above, we deem that our approximations yield acceptable results for BH formation time and gravitational mass loss from neutrino emission in CCSNe, at least for pre-SN progenitor stars that collapse into a BH within a few seconds. 
For PNSs that take much longer than one second to collapse into a BH we expect that our model will be even more accurate, since most of the PNS gravitational mass loss occurs when the PNS is relatively cold and, thus, when an approximately linear relationship exists between change in baryonic mass and gravitational mass or PNS binding energy. 
In fact, for all EOSs considered in this work $dE_{\rm{bind}}/dM_{\rm{baryon}}\simeq0.30-0.35$.

Hence, we can use our model to study EOS effects in failed CCSNe for different progenitors without having to resort to simulations that include complex detailed and computationally expensive neutrino transport. 
As discussed above, this should work even for very low-compactness progenitors, which may take weeks after core bounce to accrete enough matter to collapse into a BH.

\subsection{Computational Considerations}
\label{ssec:comp}

Due to the simplicity of our PNS model we are able to evolve a progenitor from core collapse until material is ejected from its surface using significantly less computational resources than are needed to evolve the same progenitor from core collapse to BH formation considering neutrino transport. 
Typically, each of our simulations costs between 32 and 128 CPU\,hours on the \textsc{Tetralith} supercomputer using 8 CPU cores.
This compares favorably to approximately 50 CPU\,hours per second of PNS evolution when using M1 neutrino-transport with \textsc{Flash} as in \citetalias{schneider:20}.
Furthermore, the computational cost is expected to be significant larger in fully relativistic simulations, such as those using \textsc{GR1D} as in \citetalias{ivanov:21}. 
This shows that even detailed non-relativistic simulations exploring detailed neutrino transport would be unfeasible with limited resources for stars that take $\gtrsim10\unit{s}$ to collapse into BHs. 
However, as \citetalias{ivanov:21} showed, detailed simulations may not be necessary to determine mass ejecta from failed CCSNe as long as the mass loss rate and timescale in the PNS are reasonably well described. 
As we have argued, our approach does just that.

\section{Results}
\label{sec:results}

\subsection{Pulse dynamics overview}
\label{ssec:pulse_dynamics}

Using the model described above we can estimate the neutrino emission from a PNS up to the point of BH formation and explore the aftermath of the BH-enveloping star. 
Due to neutrino emission a pressure wave appears deep within the star and propagates outwards. 
We now discuss the dynamics of this pulse. 
To facilitate direct comparisons to the work of \citetalias{fernandez:18}, we first describe the same progenitors plotted in their Figures 4 and 5, \ie a red supergiant (RSG) with $M_{\rm ZAMS}=15\,M_\odot$ (\texttt{R15z00}), a blue supergiant (BSG) with $M_{\rm ZAMS}=25\,M_\odot$ (\texttt{B25z00}), and a Wolf-Rayet (WR) star with $M_{\rm ZAMS}=40\,M_\odot$ (\texttt{W40z00}), all with solar metallicity. 
EOS effects are gauged by comparing $M_\rgrav^\rmax=2.0\,M_\odot$, proxy for a soft EOS, to $M_\rgrav^\rmax=2.5\,M_\odot$, proxy for a stiff EOS. 
While the $2.0\,M_\odot$ limit is likely a physical lower limit for BH formation from PNS collapse, $M_\rgrav^\rmax=2.5\,M_\odot$ should be close to the upper limit for BH formation and is the same limit set by \citetalias{fernandez:18}.

\begin{figure*}[htb]
\centering
\includegraphics[trim=1.82cm 0 0 0, width=0.28\textwidth]{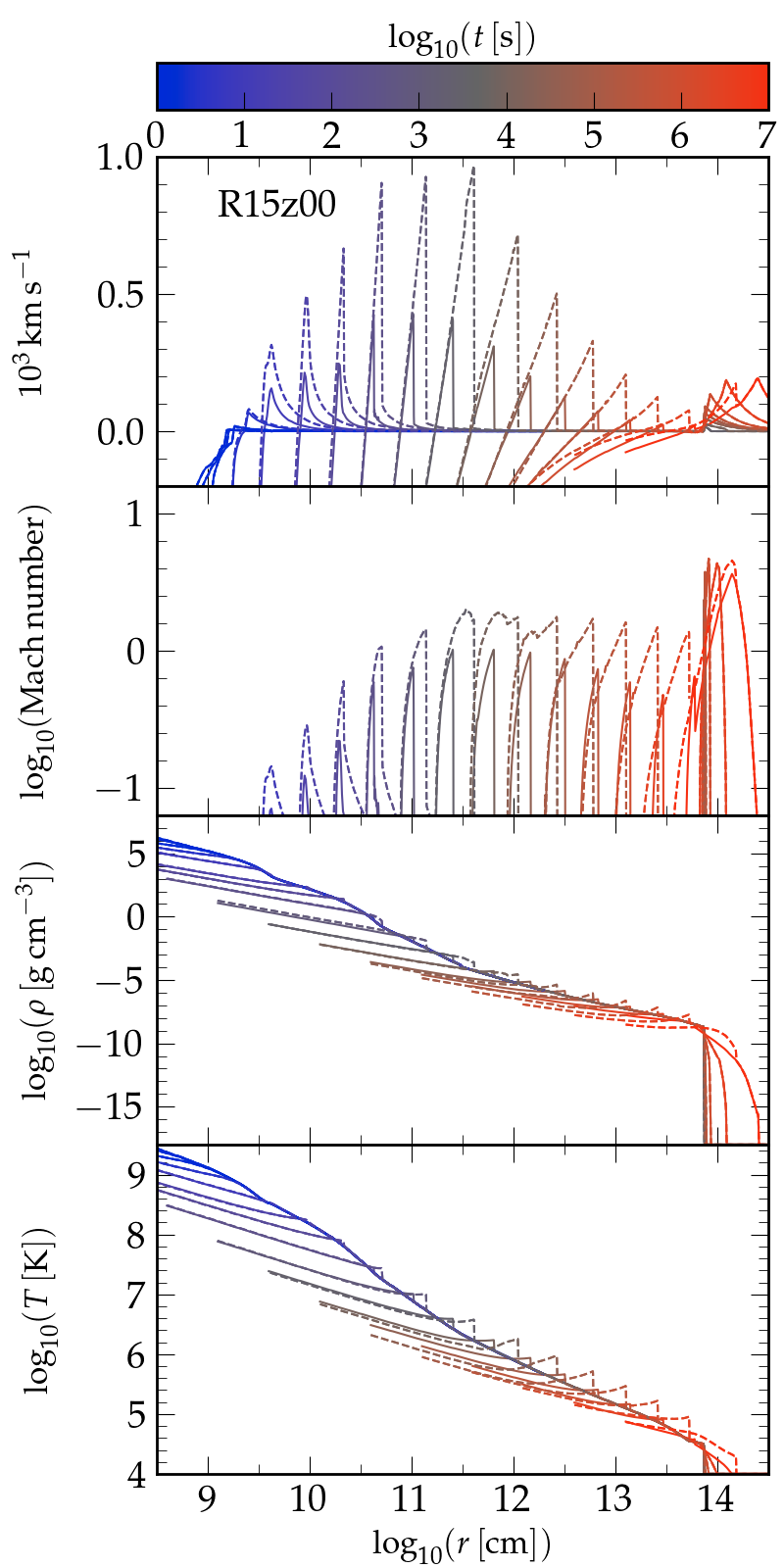}
\includegraphics[trim=1.82cm 0 0 0, clip, width=0.28\textwidth]{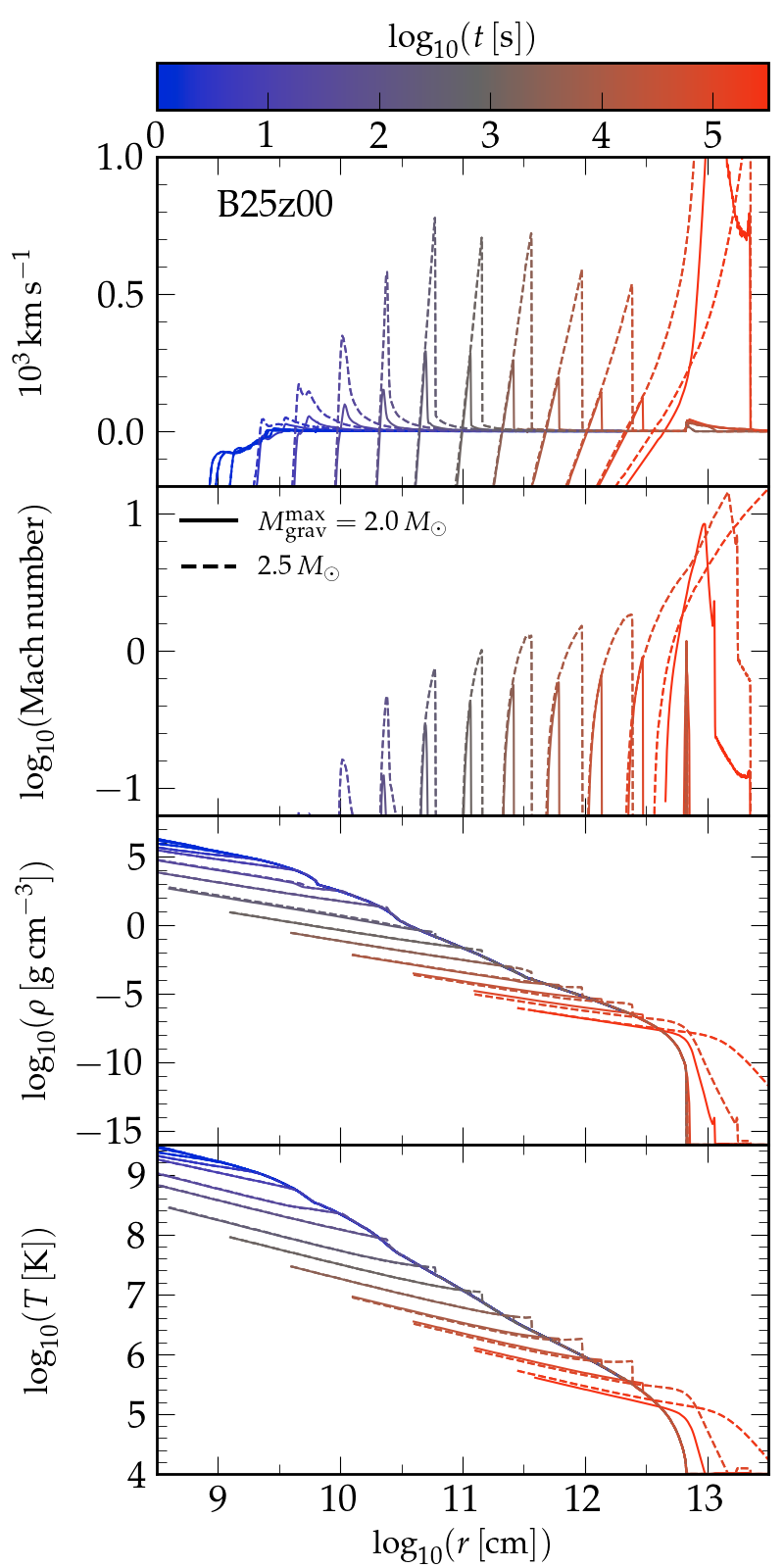}
\includegraphics[trim=1.82cm 0 0 0, clip, width=0.28\textwidth]{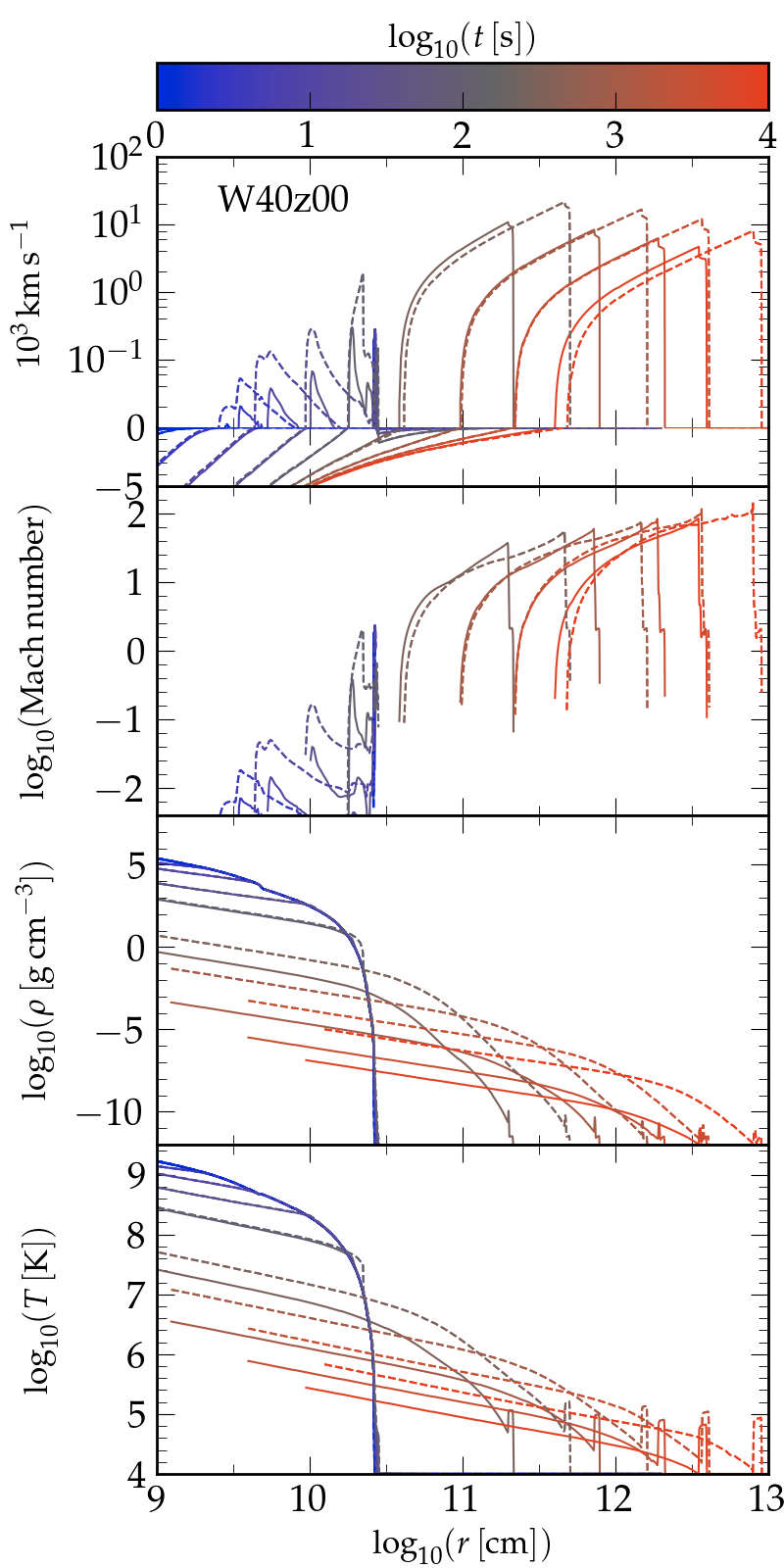}
\caption{ \label{fig:profiles} Pulse profiles as they propagate through the star at $t=10^{k/2}\unit{s}$ for $k\in\mathbb{N}$. We plot velocity (first row), Mach number (second row), density (third row), and temperature (fourth row) profiles for three pre SN progenitors, \texttt{R15z00} (left column), \texttt{B25z00} (center column), and \texttt{W40z00} (right column) for two distinct maximum mass limits $M_\rgrav^\rmax$ supported by PNSs: $M_\rgrav^\rmax=2.0\,M_\odot$ (dashed lines) and $2.5\,M_\odot$ (solid lines). 
For the \texttt{W40z00} progenitor velocity profiles we change to a logarithm scale at $10\unit{km\,s}^{-1}$.} 
\end{figure*}

In Figure~\ref{fig:profiles}, we plot stellar profiles for velocity, mach number, density, and temperature from core bounce until the simulations are stopped. 
Qualitatively and, to a degree, quantitatively our results agree well with those of \citetalias{fernandez:18}. 
For the progenitors shown in Figure~\ref{fig:profiles}, a sound pulse is created deep inside the star and first acquires a positive velocity in the carbon-oxygen shell, which extend from $10^8\unit{cm} \lesssim r\lesssim3\times10^{10}\unit{cm}$. 
The pulse propagates outwards and speeds up or slows down according to local stellar structure, accelerating to $v^\rmax \simeq 500 - 1\,000\unit{km\,s}^{-1}$ ($v^\rmax\simeq100-300\unit{km\,s}^{-1}$) for PNSs that collapse into a BH when the PNS mass reaches $M_{\rgrav}=2.5\,M_\odot$ ($2.0\,M_\odot$). 
As this occurs, pulse speed becomes comparable to local sound speed, Mach number ${\rm Ma}\gtrsim0.1$, and the pulse develops clearly defined leading and trailing edges \citepalias{fernandez:18}.

WR stars, here represented by the \texttt{W40z00} pre-SN progenitor (right plots of Figure~\ref{fig:profiles}), have their carbon-oxygen core exposed as they lose both their hydrogen and helium envelopes prior to core collapse. 
Thus, due to the relatively small size of these stars and the sharp density gradient near the stellar surface, the pulse accelerates quickly and its speed increases $\gtrsim10$ fold, reaching $\simeq100\,\unit{Ma}$ as it leaves the star.

Meanwhile, BSG stars (represented by the \texttt{B25z00} progenitor in the center plots of Figure~\ref{fig:profiles}) still retain most of their helium envelopes by core collapse and the pulse approximately maintains its speed as it propagates through this layer. 
More generally, as a sound pulse propagates through the helium envelope of a star its mass and velocity can either increase or decrease by a factor of a few, depending on the EOS and progenitor structure, before speeding up significantly as it crosses the surface of the star and reaching speeds $\simeq10\unit{Ma}$.

RSG stars (represented by the \texttt{R15z00} progenitor in the left plots of Figure~\ref{fig:profiles}), on the other hand, still maintain both their helium and hydrogen shells until core collapse. 
In these progenitors, the propagating pressure wave can loose up to $90\%$ of its velocity as it crosses the outermost layers of the star before picking up speed again and crossing the stellar surface at a few $\unit{Ma}$.

Simulations using the soft and stiff EOS proxies differ in that for stiff EOSs the pulse is both faster and more extend, indicating that it carries away more mass and energy. 
Besides the pulse formed deep within the star, we also note that a second pulse appears in the stellar surface-atmosphere interface prior to the pulse arrival. 
These surface pulses were predicted by \citet{coughlin:18} and also appear in the simulations of \citetalias{ivanov:21}, although it is difficult to estimate their significance as the atmosphere thermodynamics are not fully consistent due to the lower temperature limit of our EOS tables, $T_{\rm low}=10^4\unit{K}$.

\subsection{Ejecta properties}
\label{ssec:ejecta}

\begin{figure*}[htb]
\centering
\includegraphics[trim=1.82cm 0 0 0, width=0.28\textwidth]{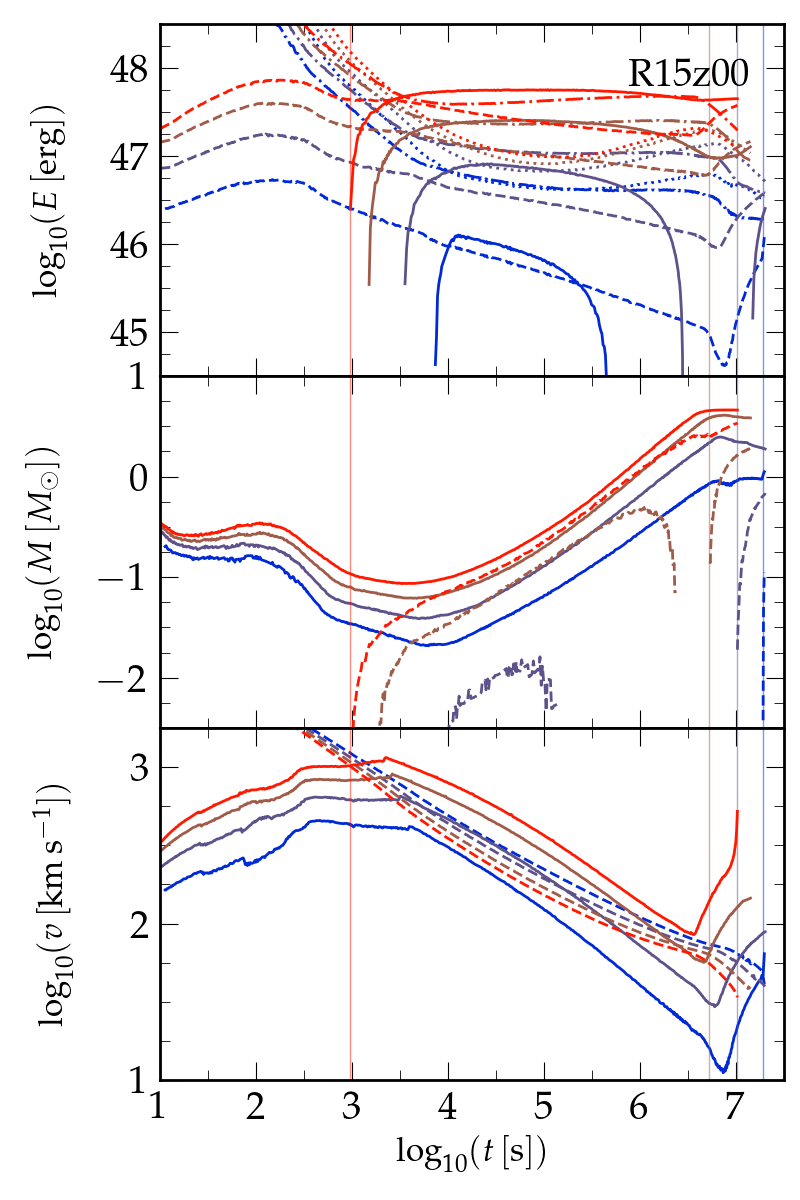}
\includegraphics[trim=1.82cm 0 0 0, clip, width=0.28\textwidth]{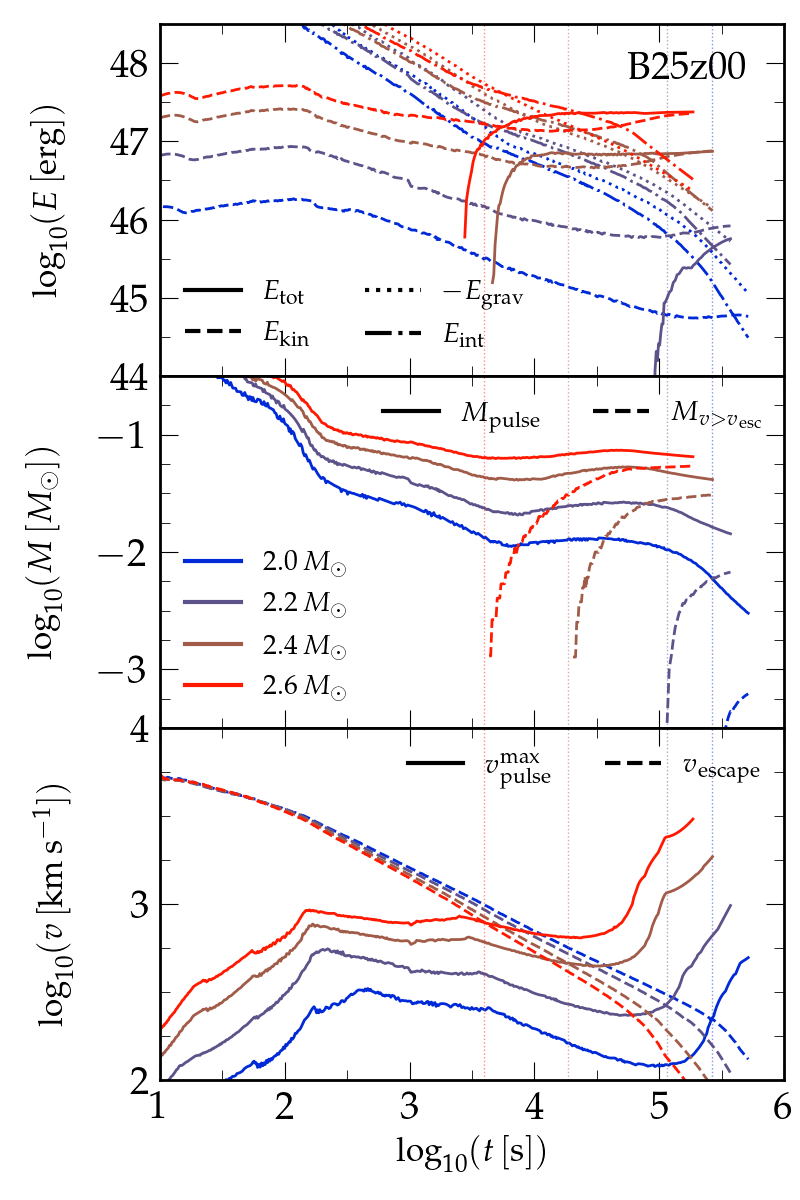}
\includegraphics[trim=1.82cm 0 0 0, clip, width=0.28\textwidth]{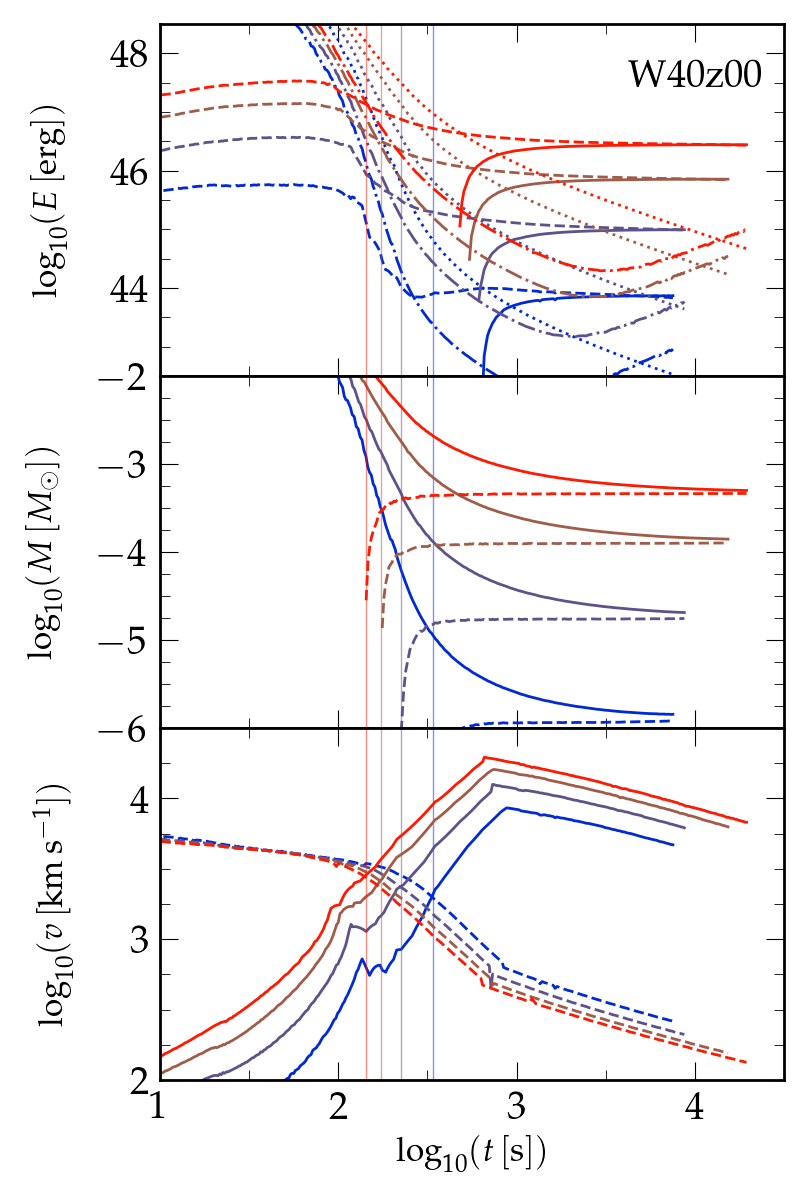}
\caption{ \label{fig:properties} Pulse properties as they propagate through the star as a function of time.  We plot internal $E_{\rm int}$, gravitational $E_{\rm grav}$, kinetic $E_{\rkin}$, and total $E_{\rtot}$ energies (first row); total pulse mass $M_{\rm pulse}$ and unbound mass $M_{v>v_{\rm esc}}$ (second row); and maximum pulse velocity $v_{\rm pulse}^\rmax$ and escape velocity $v_{\rm escape}$ where $v=v_{\rm pulse}^\rmax$. 
Plots are shown for 4 values of $M_\rgrav^\rmax$ supported by PNSs: $M_\rgrav^\rmax=2.0\,M_\odot$,$2.2\,M_\odot$, $2.4\,M_\odot$, and $2.6\,M_\odot$. 
Vertical lines mark the last moment where the maximum velocity in the pulse is smaller than the escape velocity.}
\end{figure*}

In Figure~\ref{fig:properties}, we plot the time evolution of pulse properties as they propagate outwards through the star for the same three progenitors discussed in Section~\ref{ssec:pulse_dynamics}. 
Here we define the pulse as the region of the star between the trailing edge, the stellar zone where velocity first becomes larger than 1\% of the maximum speed, and the front edge, the furthest radial coordinate that has velocity over half of the maximum speed\footnote{We exclude the pulses that develop at the stellar surface-atmosphere interface by setting a cap on the lowest density considered as part of the pulse to be at least half of the lowest density in the pulse in the previous step. We evaluate pulses starting at 10\unit{s} after core bounce and increase the analysis time by $\sim1-2\%$ every analysis step until any part of the pulse has $T<1.01\times10^4\unit{K}$ or the pulse leaves the simulation volume.}. 
This is a good approximation when the shock has developed, although it underestimates the size of the pulse early on, when the front edge is not as well defined. 
The properties are shown for four distinct values of the maximum mass supported by PNSs at the time of BH formation: $M_\rgrav^\rmax/M_\odot=2.0$, $2.2$,  $2.4$, and $2.6$. 
The $M_\rgrav^\rmax$ limits were chosen based on the most physically sound EOSs available that were still close to the bounds of our neutrino emission model.

In the first row of Figure~\ref{fig:properties}, we plot the evolution of internal energy $E_{\rm int}$, gravitational energy $E_{\rm grav}$, kinetic energy $E_{\rkin}$, and total energy $E_{\rtot}$ contained in the pulse. 
Depending on the EOS proxy, \ie the value of $M_\rgrav^\rmax$, the maximum kinetic energy imparted to the pulse due to loss of core gravitational mass changes by a factor of $\sim10$ to $\sim30$ for a given CCSN progenitor. 
Although the variation in $E_{\rkin}$ depends on progenitor, it is a monotonically increasing function of the PNS mass $M_\rgrav^\rmax$ at BH formation time. 
Also, the EOS dependent variations in $E_{\rkin}$ lead to pronounced differences in pulse properties that persist as the pulse crosses the surface of the star and part of it becomes unbound. 
The choice of EOS impacts the kinetic energy of the ejecta by as much as factor of $\sim20$ for the RSG and $\sim40$ for the BSG and WR progenitors.

In the second row of Figure~\ref{fig:properties}, we plot the evolution of the total pulse mass $M_{\rm pulse}$ and the total unbound mass $M_{v > v_{\rm esc}}$. 
The latter quantity is obtained comparing the \textit{local} pulse and escape velocities. 
We expect the true total unbound mass to be located between these two extrema as the pulse mass is increasing and the unbound mass decreasing by the end of the simulations. 
Simulations that employed softer EOSs, \ie smaller $M_\rgrav^\rmax$, showed a larger gap between $M_{\rm pulse}$ and $M_{v > v_{\rm esc}}$ at the end of the runs than simulations that used stiffer EOSs, larger $M_\rgrav^\rmax$. 
For example, $M_{\rm pulse}$ is as large as a factor of 10 than $M_{v > v_{\rm esc}}$ for simulations of the \texttt{R15z00} progenitor employing $M_\rgrav^\rmax=2.0\,M_\odot$, since this run ends at about the same time that the leading edge of the pulse becomes unbound. 
The difference is much smaller for simulations of the \texttt{W40z00} progenitor since the pulse can propagate for a long time outside of the star before simulations are terminated, allowing the necessary time for $M_{\rm pulse}$ and $M_{v > v_{\rm esc}}$ to converge. 
The unbound mass by the end of the simulation is at least $0.1\,M_\odot$ for the \texttt{R15z00} progenitor, but may be as high as $5\,M_\odot$ when considering the stiffest EOSs in our runs. 
Meanwhile, a progenitor such as \texttt{B25z00} (\texttt{W40z00}) is not likely to eject more than $0.1\,M_\odot$ ($10^{-3}\,M_\odot$), but ejects at least $\simeq10^{-3}\,M_\odot$ ($\simeq10^{-6}\,M_\odot$) through the failed CCSN mechanism. 
\citet{lovegrove:13, fernandez:18, ivanov:21} pointed out that a significant uncertainty in the mass ejected and its energy is due to pre-SN progenitor structure, while \citetalias{ivanov:21} showed that EOS effects may lead to a factor of a few uncertainty in the total mass ejected during a failed CCSN. 
However, our results show that this uncertainty may be even larger and span a few orders of magnitude for some progenitors.

Finally, we discuss the third row of Figure~\ref{fig:properties}. 
We note that the \textit{qualitative} behavior of the maximum velocity and the escape velocity where speed is at a maximum is the same for every progenitor, regardless of EOS. 
Quantitatively, however, the highest pulse velocities for each progenitor occur for EOSs that take the longest to form BHs. 
Since high outgoing velocities mean that the pulse travels faster throughout the star and, furthermore, the local escape velocity decreases with the distance to the center of the star, this means that the more time a BH takes to form the more mass becomes unbound and the earlier this happens. 
We find no cases where an increase in time to form a BH results in increased ejecta mass and lower ejecta velocity or vice versa.

\begin{table}[hbt!]
\centering
\caption{\label{tab:comp} Comparison between pulse mass $M_{\rm pulse}$, total pulse energy $E_{\rtot}$, BH formation time $t_\rBH$, and PNS gravitational mass loss $\delta M_{\rgrav}$ for the \texttt{R15z00}, \texttt{B25z00}, and \texttt{W40z00} progenitors from runs of Model I of \citetalias{ivanov:21} using the SFHo, LS220 and DD2 EOSs, the high resolution model eHR of \citetalias{fernandez:18}, and our proxy EOSs with $M_\rgrav^\rmax=2.0$ to $2.6\,M_\odot$.}
\begin{tabular}{lDDDD}
\hline
\hline
{} &
\multicolumn2c{$M_{\rm pulse}$} &
\multicolumn2c{$E_{\rm tot}$} &
\multicolumn2c{$t_\rBH$} &
\multicolumn2c{$\delta M_{\rgrav}$} \\
\hline
\multicolumn1l{\texttt{R15z00}}&
\multicolumn2c{$[M_\odot]$} & 
\multicolumn2c{$[10^{47}\unit{erg}]$} &
\multicolumn2c{$[s]$} &
\multicolumn2c{$[M_\odot]$} \\
\hline
\decimals
{SFHo}  & 2.19 & -0.119 &   2.836  &   0.196 \\
{LS220} & 2.42 & -0.103 &   2.947  &   0.222 \\
{DD2}   & 3.37 &  0.489 &$>$4.359  &$>$0.262 \\
{eHR}   & 4.2  &  1.9   &   6.1    &   0.30  \\
{2.0}   & 1.14 & -0.024 &   1.679  &   0.120 \\
{2.1}   & 1.38 &  0.041 &   2.356  &   0.157 \\
{2.2}   & 1.89 &  0.204 &   3.146  &   0.200 \\
{2.3}   & 2.72 &  0.689 &   4.061  &   0.250 \\
{2.4}   & 3.84 &  1.05  &   5.099  &   0.304 \\
{2.5}   & 4.39 &  1.88  &   6.281  &   0.361 \\
{2.6}   & 4.58 &  3.31  &   7.652  &   0.425 \\
\hline
\multicolumn1l{\texttt{B25z00}}&
\multicolumn2c{$[10^{-2}\,M_\odot]$} & 
\multicolumn2c{$[10^{47}\unit{erg}]$} &
\multicolumn2c{$[s]$} &
\multicolumn2c{$[M_\odot]$} \\
\hline
\decimals
{SFHo}  & 2.80 &  0.399 &   1.791  &   0.173 \\
{LS220} & 3.18 &  0.593 &   1.864  &   0.198 \\
{DD2}   & 5.45 &  1.76  &   2.895  &   0.261 \\
{eHR}   & 4.9  &  1.6   &   3.1    &   0.24  \\
{2.0}   & 0.30 & -0.005 &   0.880  &   0.087 \\
{2.1}   & 0.77 & -0.001 &   1.212  &   0.114 \\
{2.2}   & 1.44 &  0.032 &   1.576  &   0.144 \\
{2.3}   & 2.79 &  0.224 &   1.983  &   0.179 \\
{2.4}   & 4.14 &  0.621 &   2.457  &   0.219 \\
{2.5}   & 5.24 &  1.12  &   2.939  &   0.259 \\
{2.6}   & 6.49 &  2.06  &   3.438  &   0.299 \\
\hline
\multicolumn1l{\texttt{W40z00}}&
\multicolumn2c{$[10^{-4}\,M_\odot]$} & 
\multicolumn2c{$[10^{47}\unit{erg}]$} &
\multicolumn2c{$[s]$} &
\multicolumn2c{$[M_\odot]$} \\
\hline
\decimals
{SFHo}  & 1.44  &  0.067 &   1.535  &   0.157 \\
{LS220} & 1.63  &  0.077 &   1.570  &   0.184 \\
{DD2}   & 6.30  &  0.326 &   2.466  &   0.242 \\
{eHR}   & 5.0   &  0.25  &   2.6    &   0.22  \\
{2.0}   & 0.014 &  0.001 &   0.501  &   0.060 \\
{2.1}   & 0.062 &  0.003 &   0.806  &   0.086 \\
{2.2}   & 0.200 &  0.008 &   1.14   &   0.115 \\
{2.3}   & 0.572 &  0.027 &   1.50   &   0.147 \\
{2.4}   & 1.40  &  0.068 &   1.89   &   0.184 \\
{2.5}   & 2.86  &  0.146 &   2.32   &   0.221 \\
{2.6}   & 5.00  &  0.267 &   2.77   &   0.260 \\
\hline
\hline
\end{tabular}
\end{table}

In Table~\ref{tab:comp} we compare some of our predictions for ejecta properties and BH formation for the three progenitors discussed above considering $M_\rgrav^\rmax=2.0$ to $2.6\,M_\odot$ in $0.1\,M_\odot$ steps. 
We also include results from the interpolation mass loss model (Model I) in Table~3 for the  \citetalias{ivanov:21} for SFHo, LS220, and DD2 EOSs and results from the high resolution exponential mass loss model, eHR, in Table~2 of \citetalias{fernandez:18}.

Quantitative understanding of our results in light of those of \citetalias{fernandez:18,ivanov:21} is difficult as we do not observe any clear trends that carry across progenitors and EOSs. 
For example, for the RSG progenitor \texttt{R15z00} we observe that setting $M^\rmax_\rgrav=2.2\,M_\odot$ produces results similar to those of \citetalias{ivanov:21} using the SFHo EOSs for pulse mass $M_{\rm pulse}$, BH formation time $t_\rBH$, and total energy emitted in neutrinos $\delta M_{\rgrav}$. 
However, the total predicted pulse energy $E_{\rtot}$ is much higher in our simulations: while we predict a pulse that is unbound ($E_{\rtot}>0$), \citetalias{ivanov:21} predicts that the pulse remains bound to the star. 
For the BSG progenitor \texttt{B25z00} the results for the SFHo EOS of \citetalias{ivanov:21} are similar to ours where $M^\rmax_\rgrav=2.3\,M_\odot$ for $M_{\rm pulse}$, $t_\rBH$, $\delta M_{\rgrav}$; however, this time \citetalias{ivanov:21} predict a pulse energy $E_\rkin$ that is almost twice that of our simulations. 
Finally, for the WR star progenitor \texttt{W40z00} the SFHo results of \citetalias{ivanov:21} agree well with ours when $M^\rmax_\rgrav=2.4\,M_\odot$, except for the BH formation time $t_\rBH$, which is $20\%$ longer in our run.

The trend of BHs formaing with larger masses for more compact progenitors is exactly what is predicted by \citetalias{schneider:20} and discused here in Section~\ref{ssec:BH formation}. 
Nevertheless, from Figure~13 of \citetalias{schneider:20}, we expected that simulations employing the LS220 EOS would collapse into a BH in a slightly shorter time than those employing the SFHo EOS, which is not what was observed by \citetalias{ivanov:21} and shown here in Table~\ref{tab:comp}. 
This discrepancy may be due to the mass offset necessary to keep consistency between the inner region of the simulations of \citetalias{ivanov:21} performed with \textsc{GR1D} and the outer regions that are computed using \textsc{Flash}; see their Equation~5 and Figure~2 where the mass loss due to neutrino emission for the LS220 EOS runs has a $0.02\,M_\odot$ shift added before core bounce. 
Finally, our results for the RSG and BSG progenitors are comparable to the eHR model of \citetalias{fernandez:18}, where BHs always form with a mass of $M^\rmax_\rgrav=2.5\,M_\odot$, except for the total energy carried away by neutrinos $\delta M_{\rgrav}$ (total energy of the ejecta $E_{\rtot}$) in the RSG (BSG) case. 
Meanwhile, for the more compact progenitor star \texttt{W40z00}, \citetalias{fernandez:18} results are more in line with our $M^\rmax_\rgrav=2.6\,M_\odot$ EOS proxy.

\begin{figure*}[htb]
\centering
\includegraphics[width=0.78\textwidth]{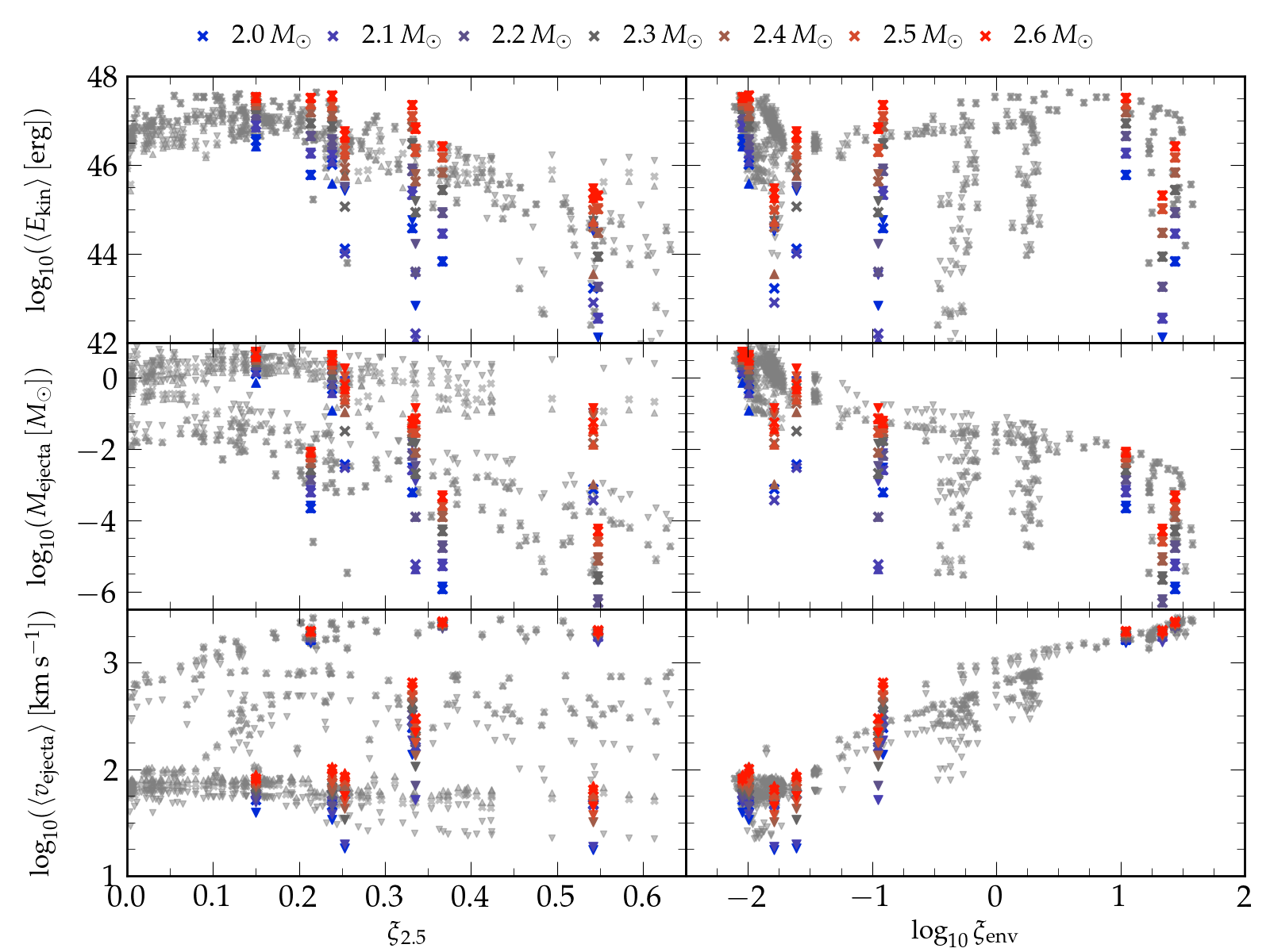}
\caption{\label{fig:F18_xi} Estimates for ejecta kinetic energy $E_{\rkin}$ (top row), mass $M_{\rejecta}$ (middle row), and average velocity $\langle{v_\rejecta}\rangle$ (bottom row) for failed CCSNe simulations for progenitors with core compactness $\xi_{2.5}$ (left column) and envelope compactness $\xi_{\rm env}$ (right column). 
Results are shown for our simulations using the \citetalias{fernandez:18} progenitors and different mass limits for BH formation between $2.0$ and $2.6\,M_\odot$ (blue to red color scale) and for another 262 progenitors found in the literature setting $M_\rgrav^\rmax = 2.5\,M_\odot$. 
Crosses indicate estimates of ejecta mass considering hydrogen recombination, while triangles pointing up consider unbound mass ignoring hydrogen recombination and triangles pointing down consider all matter in the pulse at the end of the simulation. 
}
\end{figure*}

Besides the progenitors discussed above, we simulate failed CCSN for the other progenitors of \citetalias{fernandez:18}, see Table~\ref{tab:presn}, incrementing $M_\rgrav^\rmax$ from $2.0$ to $2.6\,M_\odot$. 
In Figure~\ref{fig:F18_xi}, we summarize our results for the distribution of total pulse kinetic energy $E_\rkin$ and mass $M_{\rejecta}$ as well as its average velocity $\langle {v_\rejecta} \rangle = \sqrt{{2E_{\rkin}}/{M_\rejecta}}$, at the time the pulse reaches the atmosphere. 
Simulations with different $M_\rgrav^\rmax$ are plotted in a blue to red color-scale. 
We do not include results for the \texttt{B80z00} progenitor because, unlike results from \citetalias{fernandez:18, ivanov:21}, none of our simulations for this progenitor produced any unbound material. 
The gray symbols in the plot are results for CCSNe simulations of 262 progenitors found in the literature setting $M_\rgrav^\rmax = 2.5\,M_\odot$ and will be discussed in more detail in Sections~\ref{ssec:xi25} and \ref{ssec:xienv}.

For the quantities plotted, we consider the entire pulse at the end of the run as an upper limit of possible ejected mass, while the mass of the pulse that is unbound is taken as a lower limit. 
We also estimate a value between these two extremes considering how much mass would be ejected if all hydrogen recombination energy were converted into kinetic energy \citep{lovegrove:13}. 
The recombination energy is computed as in \citetalias{fernandez:18, ivanov:21} 
\begin{equation}
 E_{\rm rec} = \frac{\chi_H}{m_p}\int_{\rm pulse}X_H(M)dM\,,
\end{equation}
where $\chi_H=13.6\unit{eV}$ is the energy emitted by a hydrogen nucleus as it binds an electron, $m_p$ is the proton mass, and the integral over the hydrogen mass fraction $X_H(M)$ is performed over the different zones that form the pulse. 
A key difference between how \citetalias{fernandez:18, ivanov:21} determine the amount of unbound mass and how we do is that we opt to compute whether any part of the pulse is locally unbound instead of determining if the entire pulse is unbound.

First we notice that for a given progenitor uncertainty in the EOS, probed by different values of $M_\rgrav^\rmax$, can lead to changes of a few orders of magnitude in the predicted pulse ejecta energy and mass. 
This is true regardless of whether we consider the whole pulse mass or only the unbound mass, and whether we include or not hydrogen recombination energy. 
These are much wider ranges than observed by \citetalias{ivanov:21} comparing the soft DD2 EOS and the stiff SFHo EOS. 
This result is not surprising given that for these two EOSs BHs should form with initial masses that differ by $\lesssim 0.4\,M_\odot$, see Figure 13 in  \citetalias{schneider:20}, a smaller range than the one we explore here. 
In most cases, though, much of the spread in the pulse mass $M_{\rejecta}$ and kinetic energy $E_{\rkin}$ are due to simulations employing the softest EOSs, which predict a short time till BH formation and the pulses formed may not even acquire enough energy to become unbound by the time it reaches the atmosphere, even when we consider the energy input from hydrogen recombination. 
Nevertheless, despite large variations in predictions of pulse unbound mass and kinetic energy, the dispersion in the average ejected material velocity $\langle{v_\rejecta}\rangle$, whenever some material is unbound, is significantly smaller. 
Note that $\langle{v_\rejecta}\rangle^2/2=\langle{E_{\rkin}}\rangle/M_{\rejecta}$ is a measure of average energy per unit mass in both the pulse and the ejected material. 
This average velocity and, similarly, the kinetic energy per unit mass are well correlated to envelope compactness $\xi_{\rm env}$, provided that we ignore results for very soft EOSs for some of the progenitors, but mostly uncorrelated with core compactness $\xi_{2.5}$.

\subsection{Effect of core compactness}
\label{ssec:xi25}

We now fix the mass at BH formation to $M_\rgrav^\rmax=2.5\,M_\odot$ and discuss how progenitor properties affect the ejecta kinetic energy, mass and average velocity. 
To obtain a clearer picture we increase the set of progenitors studied so that we explore thoroughly the range of core and envelope compactness of pre-SN progenitors found in the literature. 
Thus, besides the 10 progenitors from \citetalias{fernandez:18}, we also simulate 52 progenitors from \citet{sukhbold:16} with solar metallicity, 36 progenitors from \citet{sukhbold:18} (12 progenitors with the standard mass loss of \cite{nieuwenhuijzen:90}, 12 with half of standard mass loss, and 12 with one tenth of standard mass loss), 126 progenitors from \citet{woosley:02} (37 with solar metallicity, 59 with ultra low metallicity, and 30 with zero metallicity), 27 rotating progenitors from \citet{woosley:06} where rotation velocities were set to zero before the start of our simulations, and 22 progenitors from \citet{laplace:21} (12 stars evolved isolated and 10 stars evolved in binary systems).

For this set of progenitors the ejecta kinetic energy $E_\rkin$ depends on core compactness $\xi_{2.5}$, top left plot of Figure~\ref{fig:F18_xi}. 
For low-compactness cores, $\xi_{2.5}\lesssim0.2$, the ejecta kinetic energy is in the range $10^{46}\unit{erg}\lesssim E_{\rkin}\lesssim10^{47.5}\unit{erg}$. 
Meanwhile, as the core-compactness increases the kinetic energy of the ejected material decreases on average.
Finally, for compact cores, $\xi_{2.5}\gtrsim0.45$, the average kinetic energy is considerably lower, $10^{42}\unit{erg}\lesssim E_\rkin \lesssim10^{46}\unit{erg}$.

The ejected mass $M_\rejecta$, however, has a different behavior than the kinetic energy $E_\rkin$. 
We observe two different ``bands'' for the ejected mass, center left plot of Figure~\ref{fig:F18_xi}. 
For one set of progenitor stars $0.1\,M_\odot\lesssim M_\rejecta\lesssim$ a few $M_\odot$ and looks independent of compactness, even for very compact progenitor cores.  
For another set of stars, though, $M_\rejecta\sim0.1\,M_\odot$ for low-compactness cores and decreases as the compactness increases. 
For this second set, progenitors with compact cores, $\xi_{2.5}\gtrsim0.5$ eject at most $\sim10^{-3}\,M_\odot$. 
As we discuss below, what distinguishes the two bands is the envelope compactness, $\xi_{\rm env}$.

Finally, the average speed of the material ejected, $\langle{v_\rejecta}\rangle$, shows three groups, which are mostly independent of core-compactness. 
In fact, the groups are well separated by mass ejected. 
For ejecta with $M_{\rejecta}\simeq1\,M_\odot$ the material has an average speed of order $30$ to $100\unit{km\,s}^{-1}$, regardless of compactness. 
Stars that eject $0.1\,M_\odot$ or less usually eject material with $300$ to $1\,000\unit{km\,s}^{-1}$, again independent of progenitor compactness. 
A third pattern is seen for stars that eject very little mass $\lesssim10^{-3}\,M_\odot$, as we will show in more detail below. 
In this case ejecta speeds exceed $2\,000\unit{km\,s}^{-1}$ except for low compactness stars, $\xi_{2.5}\lesssim0.1$, where speeds are $500-1\,000\unit{km\,s}^{-1}$.

\subsection{Effect of envelope compactness}
\label{ssec:xienv}

Low compactness pre-SN progenitors, $\xi_{\rm env}\lesssim0.1$, such as RSGs and YSGs can produce a lot of ejecta, from $\sim0.1\,M_\odot$ up to a few $M_\odot$, although this amount can be much smaller or non-existent for certain combinations of soft EOSs and progenitors. 
Whenever there is ejecta, however, it travels somewhat slowly, $\langle {v_\rejecta} \rangle \simeq 10$ to $100\unit{km\,s}^{-1}$. 
Thus, if most BHs form with masses close to that of the maximum currently known mass of a cold-NS, $M_\rgrav^\rmax\simeq2.0-2.2\,M_\odot$ \citep{antoniadis:13, cromartie:20}, then failed SNe from RSGs probably do not lead to detectable electromagnetic signatures, as their ejecta are slow and have low energy. 
Nevertheless, these systems could still shine if a disk forms due to progenitor rotation and power other types of transients \citep{woosley:12}.

Failed CCSNe of medium-compactness pre-SN progenitors, $\xi_{\rm env}\sim0.1-1$, a range that includes BSGs, often unbind $\gtrsim0.01$ to $0.1\,M_\odot$ of material. 
Although, again, certain combinations of soft EOSs and progenitors result in lower ejecta mass or even no ejecta at all. 
Qualitatively similar variations to the RSG case are seen in both the energy of the ejecta and their average velocity, which usually range from $\langle{v_\rejecta}\rangle\simeq100$ to $1\,000\unit{km\,s}^{-1}$.

Progenitors with very compact envelopes, such as WR stars, always eject some of their outer layers in the EOS limits we tested, albeit never more than $0.01\,M_\odot$ according to our simulations. 
Furthermore, for a given progenitor the ejecta always has similar average velocity in the range $\langle{v_\rejecta}\rangle\simeq1500$ to $2500\unit{km\,s}^{-1}$, regardless of the EOS and how much mass is ejected. 
If the expected ejecta velocity for this class of progenitors is indeed almost EOS independent, this would help constrain the scope of surveys similar to the one proposed by \citet{tsuna:21} to search for SNRs of failed CCSNe in the LMC.

\begin{figure}[!htbp]
\centering
\includegraphics[trim=0 2.7cm 0 0, clip, width=0.46\textwidth]{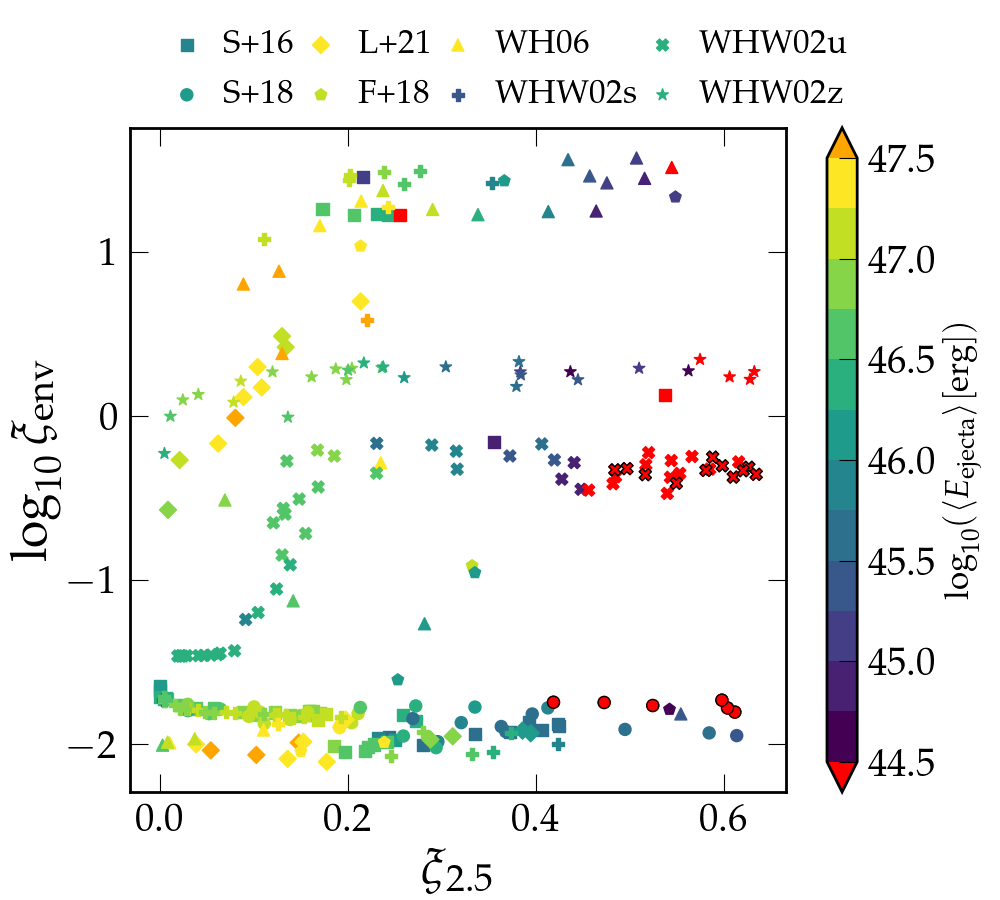}
\includegraphics[trim=0 3cm 0 0, clip, width=0.46\textwidth]{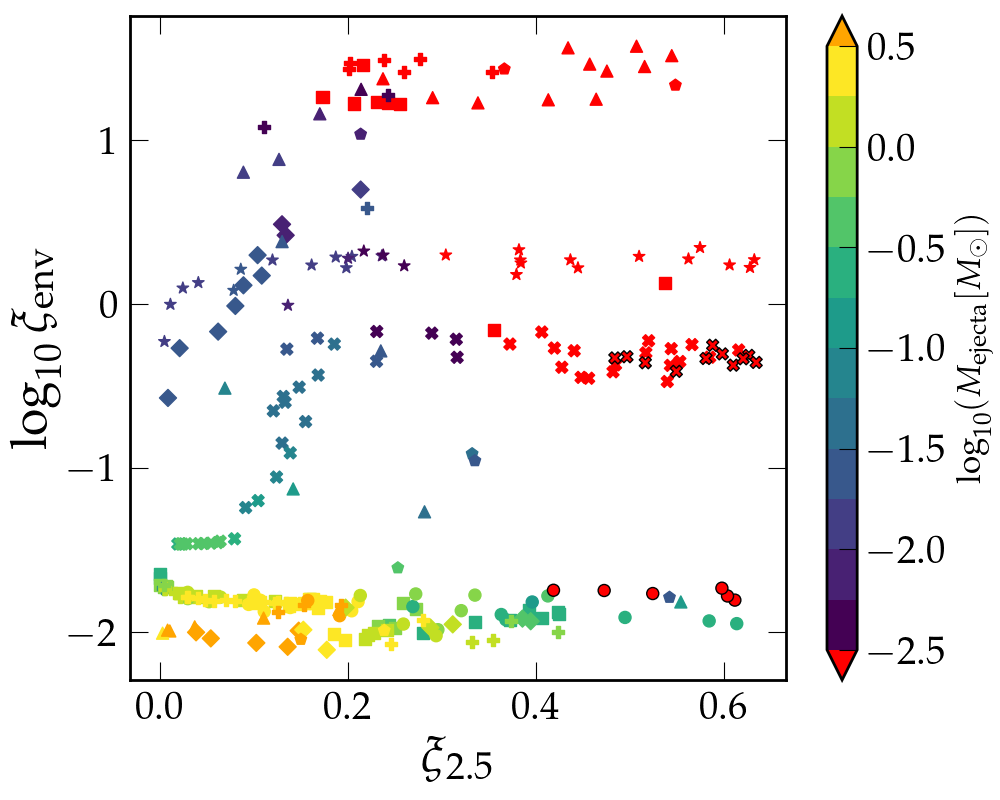}
\includegraphics[trim=0 0 0 0,       width=0.46\textwidth]{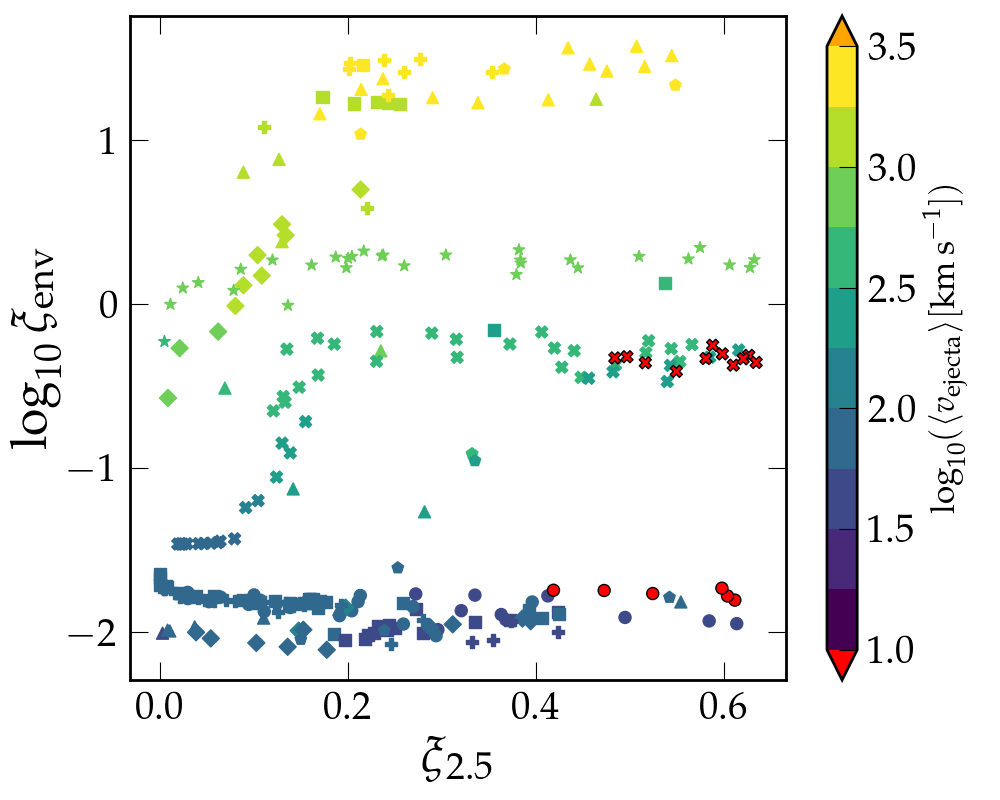}
\caption{ \label{fig:xi_ejecta} Kinetic energy (top), mass (center), and average velocity (bottom) of the ejecta as a function of core-compactness $\xi_{2.5}$ and envelope compactness $\xi_{\rm env}$ for progenitors simulated setting $M_\rgrav^\rmax=2.5\,M_\odot$.
S+16 are progenitors from \citet{sukhbold:16}, S+18 from \citet{sukhbold:18}, L+21 from \citet{laplace:21}, F+18 from \citetalias{fernandez:18}, WH06 from \citet{woosley:06}, and WHW02 from \citet{woosley:02} with s, u, and z representing solar, ultra-low ($10^{-4}z_\odot$), and zero metallicity. 
Red symbols with black contours represent failed CCSNe that did not eject any material.}
\end{figure}

In Figure~\ref{fig:xi_ejecta}, we couple the plots in Figure~\ref{fig:F18_xi} for the PNSs that form BH with $M_\rgrav^\rmax = 2.5\,M_\odot$ so that the trends just discussed become more obvious. 
We plot the kinetic energy of the ejecta, its mass, and its average speed.  The results shown are always for matter that is locally unbound when accounting for the energy available due to hydrogen recombination. 
We first notice that some regions of the $\xi_{2.5}-\xi_{\rm env}$ parameter space hardly contain any progenitors at all, while others have many progenitors. 
Specifically, the progenitors are more likely to cluster according to whether or not they retain their hydrogen ($\xi_{\rm env}\sim-2$ and often type II SN progenitors), helium ($\xi_{\rm env}\sim0$ and often type Ib SN progenitors), and carbon-oxygen ($\xi_{\rm env}\sim1.3$ and often type Ic SN progenitors) envelopes. 
Envelope properties depend mostly on ZAMS mass and metallicity and, by core collapse, are mostly independent of core properties. 
The exception is that pre-SN progenitor stars that form extended cores, $\xi \lesssim 0.2$, hardly ever have their carbon-oxygen envelopes exposed, $\xi_{\rm env}\gtrsim1$.

As discussed above, stars with extended cores and extended envelopes generally eject more mass, although this is not always the case. 
As a matter of fact, it is clear from Figure~\ref{fig:xi_ejecta} that using only the core and envelope compactness parameters $\xi_{2.5}$ and $\xi_{\rm env}$ is not enough to determine the range of possible outcomes of a failed collapse event, even when considering only a single EOS proxy. 
For example, for $\xi_{\rm env}\sim0.01$ and $\xi_{2.5}\sim0.5$ to $0.6$ we find both progenitor stars that manage to eject a few $0.1\,M_\odot$ at slow speeds as well as stars that fail to unbind any material at all. 
A similar statement is true for progenitors with intermediate envelope compactness, $\xi_{\rm env}\sim0.3$, and compact cores, $\xi_{2.5}\gtrsim0.5$: while some progenitors do eject fast material, $\langle{v_\rejecta}\rangle\simeq1\,000\unit{km\,s}^{-1}$, others do not produce any ejecta.

\begin{figure}[htb]
\centering
\includegraphics[trim=0 0 0 0, clip, width=0.46\textwidth]{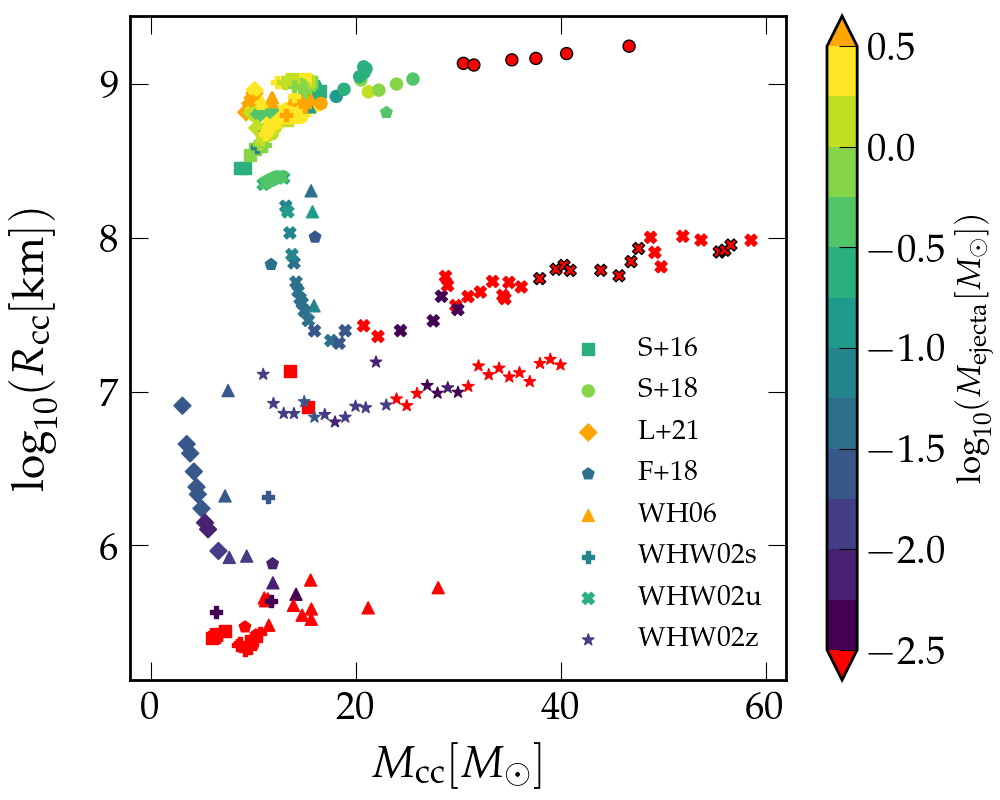}
\includegraphics[trim=0 0 0 0, clip, width=0.46\textwidth]{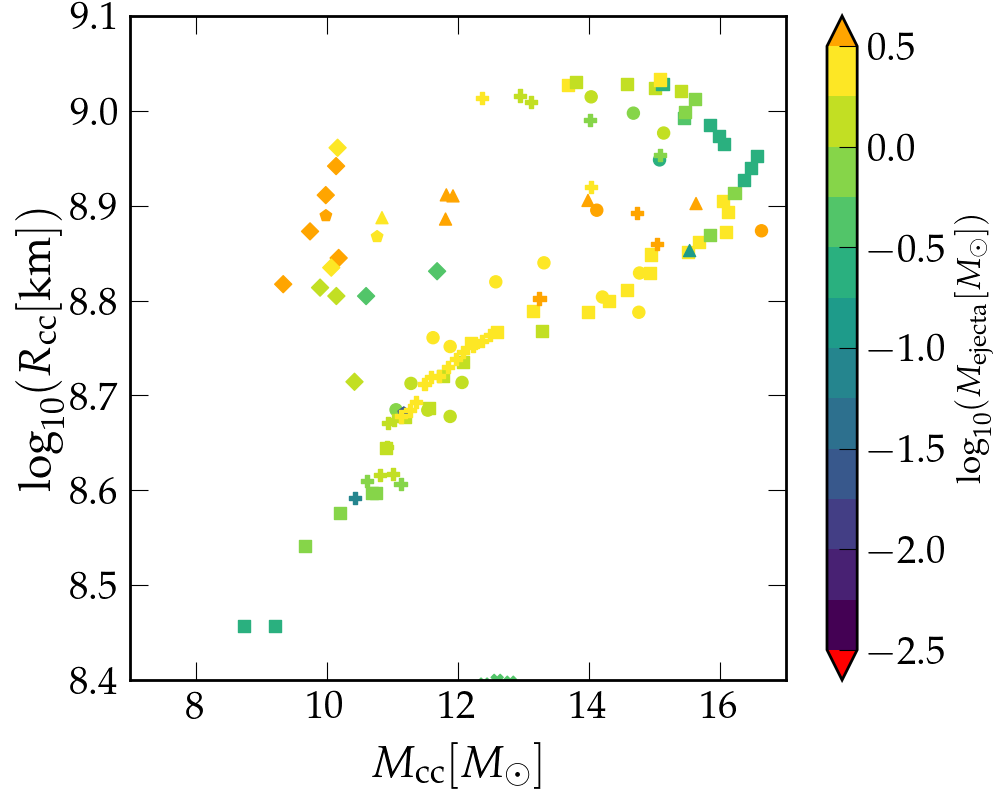}
\caption{\label{fig:MR_ejecta} Ejecta mass as a function of pre-SN progenitor mass $M_{cc}$ and radius $R_{cc}$ at core-collapse. Bottom plot is a close-up of the low-mass large-radius region where most RSGs and YSGs cluster.
Symbols are the same as in Figure~\ref{fig:xi_ejecta}.}
\end{figure}

In Figure~\ref{fig:MR_ejecta}, we show the unbound mass as a function of the progenitor mass $M_{\rm cc}$ and radius $R_{\rm cc}$ at the start of core collapse. 
Although these quantities are combined to determine envelope compactness $\xi_{\rm env}$ it is clear that they contribute separately to whether material is ejected or not. 
For example, we notice that progenitors that appear clustered in the region $\xi_{2.5}\gtrsim0.5$ and $\xi_{\rm env}\simeq0.01$ in the plots of Figure~\ref{fig:xi_ejecta} are now well separated by masses in the range from $15\,M_\odot$ to $45\,M_\odot$ and a large radius $R_{\rm cc}\simeq10^9\unit{km}$. 
Finally, from this plot we also observe that progenitors that can unbind more material through the failed-SN mechanism are those that have a lower mass as their core starts to collapse and a large radius, such as RSGs and YSGs.

\subsection{Effect of black hole formation time}

\begin{figure*}[htb]
\centering
\includegraphics[width=0.78\textwidth]{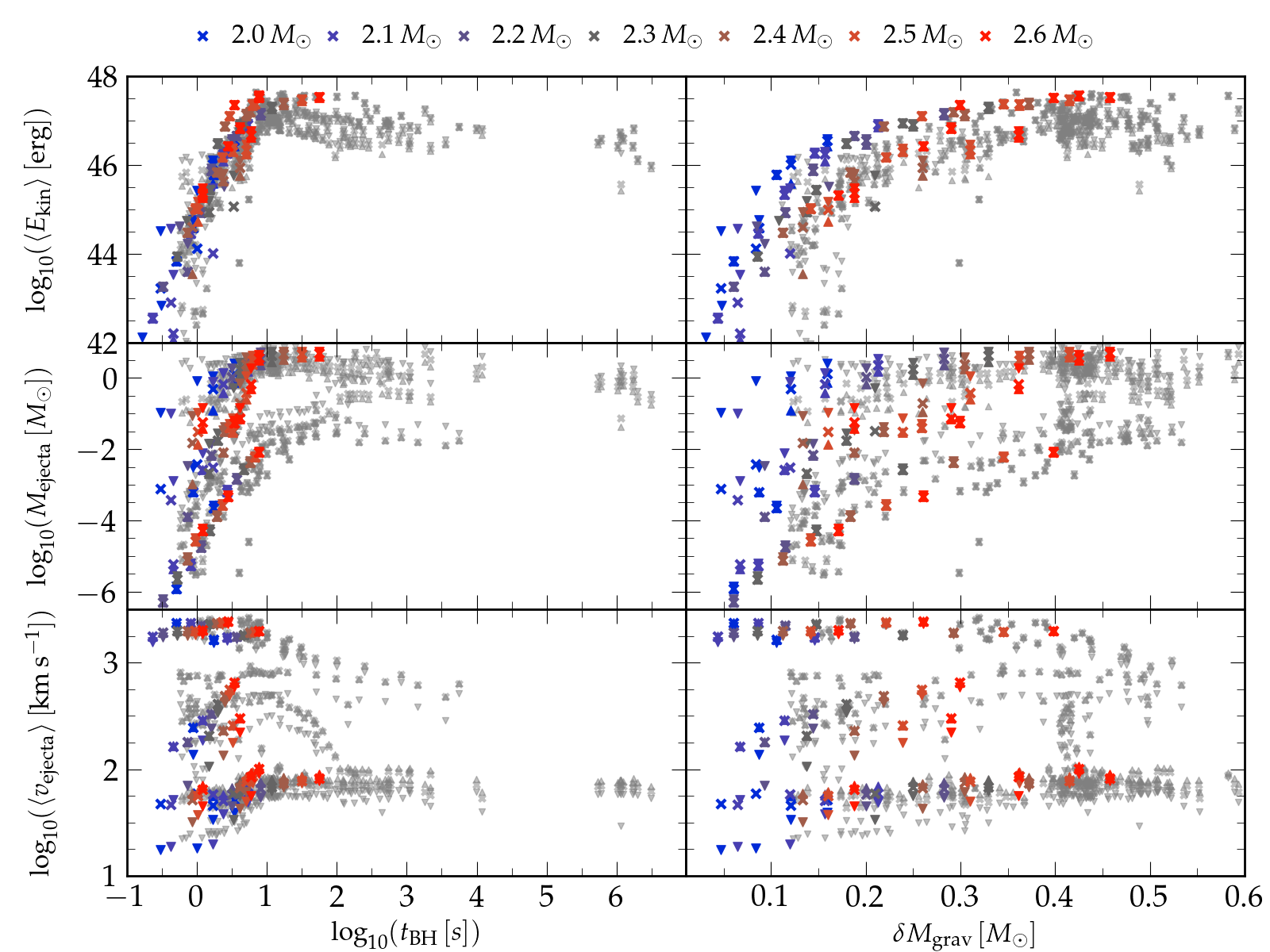}
\caption{\label{fig:F18_nu} Same as Figure~\ref{fig:F18_xi} but quantities are plotted as a function of BH formation time $t_{\rm BH}$ (left) and PNS gravitational mass loss due to neutrino emission $\delta M_{\rgrav}$ (right).
}
\end{figure*}

In Figure~\ref{fig:F18_nu} we plot the same quantities as in Figure~\ref{fig:F18_xi}, but now as a function of BH formation time $t_{\rm BH}$ and the total gravitational mass loss due to neutrino emission $\delta M_{\rgrav}$.

In the top-left plot of Figure~\ref{fig:F18_nu} we see that there is an overall scaling in the energy imparted to the pulse due to gravitational mass loss in the PNS with BH formation time that saturates if a BH takes too long to form, $t\gtrsim10\unit{s}$, as is the case for low core compactness progenitors. 
Therefore, an observational estimate of the kinetic energy of the ejecta of a failed CCSN would allow us to place some constraints on the time between the start of core collapse and BH formation. 
Such observation on the Milky Way or a nearby galaxy, where neutrino signals will also be measured \citep{sumiyoshi:06, oconnor:11}, could be used to impose further constraints on the EOS.

In the center-left plot we show the mass of the ejecta and notice two distinct trends for the ejecta, both which saturate if collapse takes $t\gtrsim10\unit{s}$. 
This is expected since the BH formation time is approximately a function of core compactness, $t_{\rBH}\propto\xi_{2.5}^{-3/2}$ \citep{oconnor:11} and, as we showed in Figure~\ref{fig:F18_xi}, that two trends emerged for the ejected or pulse mass as a function of the core-compactness $\xi_{2.5}$. 
In fact, the two trends are separated by whether or not the progenitor still has its hydrogen envelope at core-collapse, as shown in the center plot of Figure~\ref{fig:xi_ejecta}. 
Similarly, for the speed of the ejecta we observe trends that still mostly depend on envelope compactness, with little dependence on BH formation time, which depends on core compactness.

\subsection{Effect of gravitational mass loss}

In the top right side plot of Figure~\ref{fig:F18_nu} we observe that, in general, ejecta kinetic energy increases with gravitational mass loss. 
However, the trend saturates for $\delta M_{\rm grav}\gtrsim0.3\,M_\odot$ and is broken for some pre-SN progenitors. 
We also observe a weak correlation between ejecta mass and total gravitational mass loss, although the scatter is a few orders of magnitude. 
Finally, for the average ejecta speed we notice three strips that are mostly uncorrelated with the PNS binding energy, 
each one related to the progenitor envelope properties. 
Thus, it is hardly possible to constrain binding energy of the PNS at the time of collapse from measurements of the velocity of the ejecta alone.

\subsection{Mass ejected in the Hertzsprung–Russell diagram}

\begin{figure}[htb]
\centering
\includegraphics[trim=0 0 0 0, clip, width=0.46\textwidth]{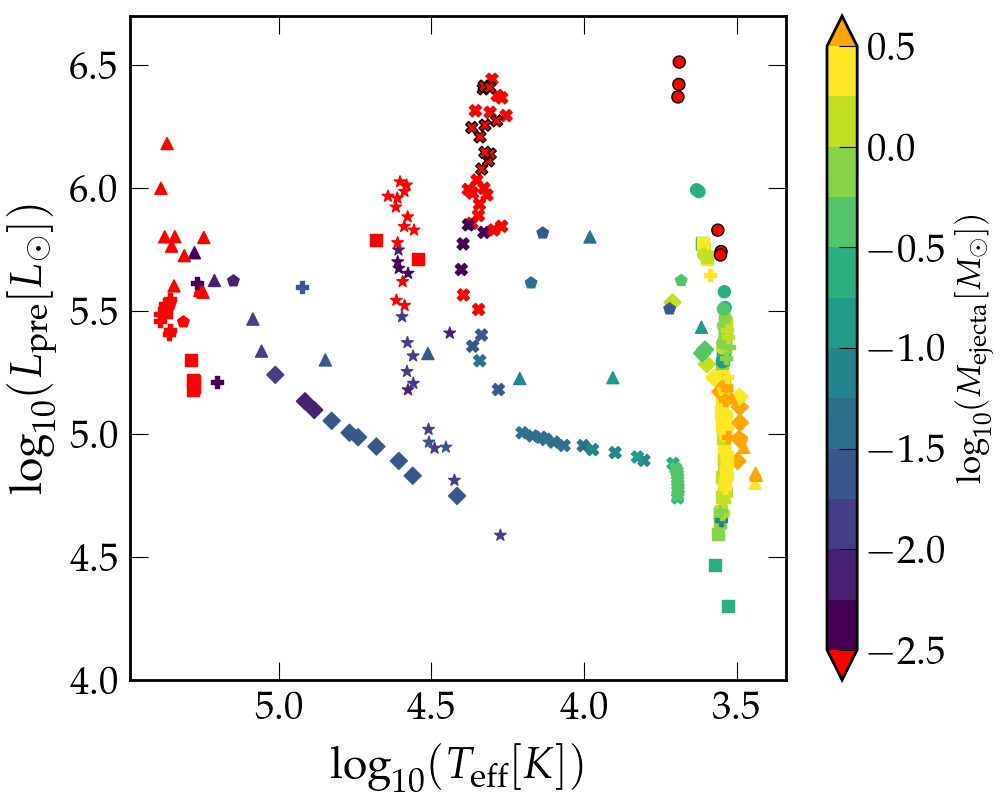}
\caption{\label{fig:HR_ejecta} Ejecta mass as a function of pre-SN progenitor luminosity $L_{\rm pre}$ and effective surface temperature $T_{\rm eff}$ at the start of core-collapse. 
Symbols are the same as in  Figure~\ref{fig:xi_ejecta}. 
Black border denote progenitors that did not eject any mass.}
\end{figure}

When observing a distant star we can measure neither its mass nor its radius directly. 
Instead, we have to rely on stellar evolution codes to infer those observables from the surface temperature and luminosity of the star. 
Here, we approximate the pre-SN stellar luminosity using the fit of \citet{sukhbold:18} 
\begin{equation}\label{eq:sukhbold}
L_{\rm pre}\simeq5.77\times10^{38}\left(\frac{M_{\rm He}}{6 M_\odot}\right)^{3/2} \unit{erg\,s}^{-1}\,
\end{equation}
where $M_{\rm He}$ is the mass bound by the Helium core. 
For stars that have lost their helium shell we use the carbon-oxygen core mass $M_{\rm CO}$. 
The effective surface temperature $T_{\rm eff}$ is estimated from the luminosity $L_{\rm pre}$ and the pre-SN progenitor radius $R_{\rm cc}$. 
For pre-SN progenitors that have their luminosity and surface temperatures available, such as those of \citetalias{fernandez:18} and \citet{laplace:21}, we obtain  deviations in the 10\% range using this approximation and opt to use it for all other stars. 
A few exceptions do occur, but deviations are at most a factor of two from the approximation of Equation~\eqref{eq:sukhbold}.

Using the approximations above, we plot the ejected mass as a function of the observable quantities $L_{\rm pre}$ and $T_{\rm eff}$ in Figure~\ref{fig:HR_ejecta}. 
The diagram shows that the hottest progenitors at core collapse, $T_{\reff}\gtrsim10^{4.2}\unit{K}$ usually collapse into a BH without a significant transient as they eject little to no mass, $M_{\rejecta}\lesssim0.1\,M_\odot$. 
These progenitors are often the evolutionary endpoints of massive stars with low or zero metallicity \citep{woosley:02}, stars in binary systems \citep{laplace:21}, or fast rotating stars that mix their shells while in the main sequence \citep{woosley:06}. 
Recall that we removed the angular momentum to simulate the latter. 
Because these progenitors have lost their hydrogen shells by core collapse time, in the case of a successful SN explosion they are observed as type Ib or type Ic SNe, stars that have their strongly bound helium or carbon-oxygen cores exposed, respectively.

Stars on the cold end of the diagram, RSGs or YSGs and $T_{\reff}\lesssim10^{3.7}\unit{K}$, are more likely to unbind their outer layers in the case of a failed SN. 
This is expected, as the hydrogen envelope these colder stars retain until their core starts to collapse are loosely bound. 
When these stars explode succesfully, they are observed as type II SNe due to their hydrogen rich ejecta. 
Progenitors with luminosity $L\lesssim10^{5.1}L_\odot$ have been associated with type II SN while, to date more luminous RSGs and YSGs have not \citep{rodriguez:22}. 
Thus, these stars are reasonable candidates to be progenitors of failed SNe.

\section{Summary \& Conclusions}
\label{sec:conclusions}

We developed simple templates to model the entropy evolution and neutrino emission rates of PNSs formed in CCSNe. 
The templates were built based on \textsc{Flash} simulations that used M1 neutrino transport for 25 selected progenitors using the baseline SRO EOS of \citetalias{schneider:20}. 
Simulations using our templates reproduce the BH formation time and the PNS binding energy at BH formation time with a relative error $\lesssim20\%$ for an intermediate compactness progenitor and different hadronic EOSs. 
The template also reproduces the PNS gravitational mass and entropy as well as the average neutrino emission rate of PNSs within a few percent.

Once built, the templates allow us to carry out the evolution of massive stars that undergo failed core collapse, \ie a supernova where the shock is not revived, beyond the moment a PNS becomes a BH and until material is ejected from the surface of the star. 
The ejecta results from the disturbance in the hydrodynamic equilibrium in the outer layers of the star due to PNS gravitational mass loss by neutrino emission \citep{nadyozhin:80, lovegrove:13}. 
Due to the simplicity of the emission model we can simulate core collapse events where the PNS takes less than a second to up to weeks to accrete enough material to collapse into a BH.

With this in mind we explored how the EOS can affect failed CCSNe ejecta for different progenitor stars. 
In our model, EOS effects are gauged by setting the BH formation at different PNS masses. 
Despite the many simplifications in the core-collapse physics, our results expand upon previous works. 
The most significant improvement being the inclusion of a wide range of pre-SN progenitor stars with different ZAMS masses and core and envelope compactness.

As \citetalias{ivanov:21}, we also observe that EOS stiffness can significantly alter the total amount of mass ejected and its average energy, sometimes by a few orders of magnitude. 
However, the average energy per unit mass, probed by the average ejecta speed, is often not considerably affected by changes in EOS. 
This is especially true for stars with compact envelopes, such as WR stars, where only a small fraction of the envelope, $M_{\rejecta}\ll10^{-1}\,M_\odot$, is ejected with speeds in excess of $1\,000\unit{km\,s}^{-1}$. 
Direct comparison of our results to those of \citetalias{fernandez:18, ivanov:21} show that changes in how the collapse is modeled can modify the predicted properties of the ejected mass in a failed core collapse in ways that are difficult to predict. 
Nevertheless, the overall qualitative picture is always the same, stiffer EOSs predict longer times to form BHs which in turn results in higher neutrino emissions and, thus, larger ejecta mass with higher average kinetic energy.

Our results also show that, even when considering that all PNSs form BHs with the same gravitational mass, core and envelope compactness, $\xi_{2.5}$ and $\xi_{\rm env}$, respectively, are not enough parameters to predict the ejecta properties of a failed CCNS. 
Although these two parameters do allow us to make educated guesses of the ejecta mass and energy for most stars, in some cases it is also necessary to consider the total mass of the star and/or its radius at the moment of core collapse separately.

An assumption made in this work and facilitated by the fact that we only perform spherically symmetric CCSNe simulations, where the SN shock is rarely revived, was that every progenitor could collapse into a BH. 
However, successful SNe are routinely observed and the pre-SN progenitor envelope properties can be inferred from the emission lines from the SN explosion. 
Sometimes, even the pre-SN progenitors themselves are identified by direct comparison of SN images to previous observations of the same region of the sky. 
In fact, pre-SN progenitors that retain their hydrogen envelopes evolve into RSGs and YSGs and have been reported to explode as type II SNe \citep{smartt:09,  williams:14, rodriguez:22}. 
At core-collapse time, these stars have an effective surface temperature $\log_{10}(T_{\reff}\,[{\rm K}])\simeq3.6$ and, although their luminosity spans the range $\log_{10}(L_{\rm pre}/L_\odot)\simeq4.5-6.5\unit{dex}$, to date only progenitors with $\log_{10}(L_{\rm pre}/L_\odot)\lesssim5.1\unit{dex}$ have been directly associated to successful SN explosions \citep{rodriguez:22}. 
Thus, many of the stars in our dataset, particularly those with $\log_{10}(T_{\reff}\,[{\rm K}])\simeq3.6$ and $\log_{10}(L_{\rm pre}/L_\odot)\gtrsim5.1\unit{dex}$, as seen in Figure~\ref{fig:HR_ejecta}, could be progenitors of failed SNe as currently suggested by observations. 
On the other hand, spherically symmetric simulations by \citet{boccioli:22a} employing neutrino-driven turbulent convection using time-dependent mixing length theory model \citep{couch:20, boccioli:22}, suggest that some low-mass low-compactness progenitors, $M_{ZAMS}\lesssim15\,M_\odot$ and $\xi_{2.5}\lesssim0.25$ and which we find to have luminosity $L\lesssim10^5\,L_\odot$, could also lead to failed CCSNe.

Another limitation of our work is that we did not include rotation. 
Angular momentum effects are very important when simulating CCSNe \citep{oconnor:11, dessart:12, richers:17} and may lead to the development of accretion disks around the PNS which may facilitate mass ejection \citep{feng:18, batta:19, murguia-berthier:20}. Thus, with respect to rotation, our results could be taken as lower limits to the amount of mass ejected and energy for a particular combination of progenitor and EOS.
Our model could be extended to add rotation, however, this would result in an added dimension to the PNS physics and its evolution and neutrino emission templates would have to be updated accordingly.

From a practical perspective, one could compare our results to those of a failed CCSNe from a slowly or non-rotating progenitor. 
Then, if a failed CCSN were observed, its progenitor identified, and total ejected mass and velocity estimated, one could, in principle, validade stellar evolution models as well as place limits on the EOS of hot-dense matter and PNS properties at BH formation time based on the trends observed in the results presented here. 
Ideally, this would be done using a more complete picture of failed core collapse; 
one built from a set of simulations including more progenitors, other EOS templates validated by multi-dimensional simulations, other EOS proxies, and including rotation.

\begin{acknowledgments}

The authors would like to thank S.~Couch for code development within FLASH and early discussions related this work. We would also like to thank R.~Fern\'{a}ndez, for his help on how to set up the stellar surface atmosphere interface, and E.~Laplace and D.~Vartanyan for help with the progenitor stars from their work. 
We are also grateful to A.~Betranhandy, S.~Zha, O.~Eggenberger Andersen, and H.~Andresen, for fruitful discussions during development and writing of this manuscript. 
The authors acknowledge support from the Swedish Research Council project No. 2020-00452.
The simulations were performed on resources provided by the Swedish National Infrastructure for Computing (SNIC) at PDC and NSC. 
Partially funded by the Swedish Research Council through grant afreement No. 2018-05973 
The software used in this work was in part developed by the DOE NNSA-ASC OASCR Flash Center at the University of Chicago. 

\software{FLASH \citep{fryxell:00, dubey:09,couch:13, oconnor:18},  
NuLib \citep{oconnor:15}, 
SROEOS \citep{schneider:17}, 
Timmes EOS \citep{timmes:99},
Matplotlib \citep{hunter:07},
NumPy \citep{harris:20},
SciPy \citep{virtanen:20},
MESA \citep{paxton:11, paxton:13, paxton:15, paxton:18, paxton:19},
VisIt \citep{HPV:VisIt}.
}

\end{acknowledgments}

%
%

\bibliography{LLSNe}

\end{document}